\documentclass[11pt]{article}
\pdfoutput=1
\usepackage{jheppub}
\usepackage[usenames,dvipsnames,table]{xcolor}
\usepackage{graphicx,amsmath,amssymb,amsthm,multirow,array,bm,bbm,esint,slashed}
\usepackage{xfrac,faktor}
\usepackage[mathscr]{eucal}
\usepackage{subfig}

\newcommand\beq{\begin{equation}}
\newcommand\eeq{\end{equation}}

\title{A master bosonization duality}

\author[a]{Kristan Jensen}
\affiliation[a]{Department of Physics and Astronomy, San Francisco State University, San Francisco, CA 94132}

\preprint{\today}

\emailAdd{kristanj@sfsu.edu}

\abstract{We conjecture a new sequence of dualities between Chern-Simons gauge theories simultaneously coupled to fundamental bosons and fermions. These dualities reduce to those proposed by Aharony when the number of bosons or fermions is zero. Our conjecture passes a number of consistency checks. These include the matching of global symmetries and consistency with level/rank duality in massive phases.}

\begin{document}
\maketitle

\section{Introduction}

One of the best understood dualities is ``level/rank'' duality in Chern-Simons (CS) gauge theories. An infinite sequence of level/rank dualities equates
\beq
	\label{E:levelrank}
	SU(N)_{-k} \quad 
	\leftrightarrow 
	\quad U(k)_N\,,
\eeq
for all $N,k\geq 1$, where the subscript indicates the Chern-Simons level. The theories are dual in that the observables of both theories are identical~\cite{Naculich:1990pa,Mlawer:1990uv,Nakanishi:1990hj}.

There has been an accumulation of evidence this decade for dualities between non-supersymmetric Chern-Simons theories coupled to fundamental matter~\cite{Aharony:2011jz,Giombi:2011kc,Aharony:2012nh,Aharony:2015mjs,Hsin:2016blu}. These conjectured dualities may be thought of as taking the level/rank dualities, adding suitable matter content on both sides, and tuning to a conformal field theory (CFT). The basic sequences of interest in this work were precisely formulated by Aharony and read~\cite{Aharony:2015mjs}:
\begin{align}
	\label{E:sequence1}
	SU(N)_{-k+\frac{N_f}{2}} \quad \text{ with }N_f \text{ Dirac fermions }\quad 
	&\leftrightarrow 
	\quad U(k)_N \quad \text{ with } N_f \, \text{ scalars}\,,
	\\
	\label{E:sequence2}
	U(k)_{N-\frac{N_f}{2}} \quad \text{with } N_f \text{ Dirac Fermions } \quad
	& \leftrightarrow
	\quad SU(N)_{-k} \quad \text{ with }N_f \text{ scalars}\,.
\end{align}
There are also time-reversed versions of the dualities. All interactions in the fermionic theories arise from the gauge interactions, while the scalars are understood to be ``Wilson-Fisher'' (WF) scalars, meaning that on the scalar side of the duality one turns on a $|\phi|^4$ potential and mass term and tunes to criticality. Both dualities require $N_f \leq k$, although there is a proposal~\cite{Komargodski:2017keh} to extend the dualities slightly beyond this ``flavor bound.'' Because the dualities~\eqref{E:sequence1} and~\eqref{E:sequence2} relate theories with fundamental fermions to theories with fundamental bosons, they have been dubbed ``3d bosonization.''

These dualities have been the subject of recent attention from a variety of viewpoints. For the special case $N=k=N_f=1$, these dualities are related to the surface states of time-reversal invariant topological insulators and the fractional quantum Hall effect at half filling~\cite{Son:2015xqa,ws,mv,mam}, lead to a web of dualities~\cite{Karch:2016sxi,Seiberg:2016gmd,Murugan:2016zal,Karch:2016aux}, and can even be proven on the lattice~\cite{Chen:2017lkr}. They are crucial actors in mapping out the phase diagram of QCD$_3$ as well as some of its cousins. At large $N, k$, with $N/k$ finite, they are dual to a peculiar theory of gravity known as Vasiliev theory~\cite{Aharony:2011jz,Giombi:2011kc,Aharony:2012nh}. These theories are in fact solvable in this limit, and much is known of their thermal physics and scattering amplitudes~\cite{Aharony:2011jz,Giombi:2011kc,Aharony:2012nh,Aharony:2012ns,Jain:2014nza,Inbasekar:2017ieo}. Away from $N=k=N_f=1$, they imply a web of dualities for gauge theories with product gauge groups and (bi)fundamental matter~\cite{Jensen:2017dso}, known as quiver gauge theories, and have been embedded intro string theory~\cite{Jensen:2017xbs} (see also~\cite{Armoni:2017jkl}). For other interesting works see e.g.~\cite{Radicevic:2015yla,Aharony:2016jvv,Metlitski:2016dht,Gaiotto:2017tne,Gomis:2017ixy,Cordova:2017vab,Aitken:2017nfd}.

The dualities~\eqref{E:sequence1} and~\eqref{E:sequence2} remain unproven, and in the absence of supersymmetry, it is difficult to envisage a proof. Nevertheless there is significant evidence that they are true. The best evidence comes from direct computations at large $N, k$ with $N/k$ finite. The exact global symmetries and their 't Hooft anomalies match~\cite{Benini:2017dus}, as do the quantum numbers of baryon and monopole operators~\cite{Aharony:2015mjs}. At large $N$, these dualities appear to be inherited from a three-dimensional version of Seiberg duality~\cite{Jain:2013gza,Gur-Ari:2015pca}, and there is some expectation that the non-supersymmetric bosonization dualities are always the offspring of a parent supersymmetric duality (see e.g.~\cite{Aharony:2012ns,Jain:2013gza,Gur-Ari:2015pca,Kachru:2016rui,Kachru:2016aon}).

Finally, it is expected that the Chern-Simons-matter theories in~\eqref{E:sequence1} and~\eqref{E:sequence2} possess at least one relevant operator, the mass operator for the fundamental fermions or bosons. Deforming by this mass operator triggers a flow to a massive phase, described in the infrared (IR) by a topological field theory (TFT). The precise low-energy TFT depends on the sign of the mass. One then thinks of the Chern-Simons-matter theories in~\eqref{E:sequence1} and~\eqref{E:sequence2} as describing a second order transition separating these two phases. Crucially, the IR TFTs describing the massive phases of one side of the duality match those of the other~\cite{Aharony:2015mjs}. For example, deforming $SU(N)_{-k+\frac{N_f}{2}}$ theory coupled to $N_f$ fermions by a negative mass leads to a $SU(N)_{-k}$ TFT in the IR, while deforming $U(k)_N$ theory coupled to scalars by a positive mass-squared leads to an IR $U(k)_N$ TFT. These TFTs are identical by virtue of level/rank duality. A similar computation matches the other phases.

In this note we propose a new infinite sequence of bosonization dualities between Chern-Simons-matter theories with both fundamental fermions and bosons.\footnote{A related conjecture was made some years ago for $N_f=N_s=1$~\cite{Jain:2013gza}.} It reads
\beq
	\label{E:proposal}
	SU(N)_{-k+\frac{N_f}{2}} \,\,\text{ with }N_f \, \psi\,, N_s \, \phi \quad
	\leftrightarrow
	\quad U(k)_{N-\frac{N_s}{2}} \,\, \text{ with }N_f\, \Phi\,, N_s\, \Psi\,,
\eeq
where $\psi$ indicates a Dirac fermion and $\phi$ a WF scalar, and we require $N_f\leq k, N_s \leq N$ but $(N_f,N_s)\neq (k,N)$. On both sides we impose a manifest $SU(N_f)\times SU(N_s)\times U(1) \times U(1)$ global symmetry. (On the $SU(N)$ side the $U(1)$'s rotate the fundamental fermions and bosons, while on the $U(k)$ side one rotates the fundamental fields but the other is a monopole symmetry.) As we will discuss, at large $N$, the only relevant or marginal operators consistent with this symmetry are scalar and fermion masses as well as $ (\bar{\psi}\cdot \phi)(\phi^{\dagger}\cdot \psi)$, where the dots indicate a contraction of gauge indices. This last operator is the unsung hero of this note. It is important for the following reason. The theories in~\eqref{E:proposal} have a phase in which the gauge group is partially Higgsed, and in this phase this operator gives a mass $\sim |\phi|^2$ to the fermions which are neutral under the unbroken gauge group. To find its dimension at large $N$, we use that at large $N$ the theories in~\eqref{E:proposal} may be obtained as the endpoint of a double trace flow from the free-field fixed point triggered by $(|\phi|^2)^2$. The quartic operator picks up a $1/N$ suppressed anomalous dimension relative to the free-field fixed point, $\Delta = 3 + O(1/N)$, so that it is approximately marginal. In order for the duality to work, we find that it must be present on both sides of the duality with a coefficient whose sign is that of the Chern-Simons level, i.e. negative for the $SU(N)_{-k+\frac{N_f}{2}}$ theory and positive for the $U(k)_{N-\frac{N_s}{2}}$ theory. More precisely, we require these signs in the deep IR of the Higgs phases. Curiously, in supersymmetric Chern-Simons-matter theories with level $k$ and at least $\mathcal{N}=2$ supersymmetry, the coupling of this operator is fixed to be $\sim 1/k$.

As far as we know the sign of the $O(1/N)$ correction to this dimension has not yet been computed. If it is positive, then this operator is slightly irrelevant yet important in the IR, in which case it is dangerously irrelevant. If the sign is negative so that the operator is relevant, then its coefficient is not a free parameter and must be tuned to realize a CFT. Both possibilities are of interest and warrant future study. In either case it is clear that the sign of this coupling in the deep IR is not a choice but must be fixed by the dynamics, and it is not clear if the dynamics choose the sign we need for the duality to hold. Turning the matter around, if we regard the duality~\eqref{E:proposal} as sacrosanct, then we are making an implicit claim about the flow of this operator which would be nice to explicitly check at large $N$.\footnote{We would like to thank O.~Aharony and Z.~Komargodski for discussions on these points.} 

We perform several basic consistency checks on our proposal. The first is to map out the phase diagram of both sides of~\eqref{E:proposal}, under the assumption that the fermion and scalar mass operators remain relevant. The theories in~\eqref{E:proposal} may be understood as a multi-critical point in which both of these masses are tuned to vanish. We argue that the ensuing two-dimensional phase diagram has five distinct phases, all visible semiclassically, described by four different IR TFTs. These phases are separated by critical lines described by the theories in~\eqref{E:sequence1} and~\eqref{E:sequence2}, and a critical line described by $N_f N_s$ free fermions and a decoupled TFT.

We also discuss the quantum numbers of baryons and monopoles in these theories, finding that, as in the basic bosonization dualities~\eqref{E:sequence1} and~\eqref{E:sequence2}, the baryons of one side may be consistently matched to monopoles in the other. Finally we deduce the exact global symmetries of both sides of~\eqref{E:proposal} and find that they match.

In the title of this note we call the proposal~\eqref{E:proposal} a ``master'' duality. In giving this presumptuous name we have two things in mind. The first is that this proposal reduces to Aharony's when $N_s=0$ or $N_f=0$. The second concerns recent works which use the basic dualities as ``seed dualities'' to generate new ones by gauging global symmetries on both sides of the seed~\cite{Karch:2016sxi,Seiberg:2016gmd,Murugan:2016zal,Karch:2016aux,Jensen:2017dso}. In the context of quiver gauge theories, one of the results of~\cite{Jensen:2017dso} is that one can dualize node-by-node: given a quiver with a $SU$ or $U$ gauge group factor coupled to only bosons or fermions, one can generate a dual quiver by replacing a node and the matter attached to it with its dual according to~\eqref{E:sequence1} and~\eqref{E:sequence2}. Assuming our conjecture~\eqref{E:proposal}, one can use the same logic to dualize any node of a quiver with fundamental matter.

The remainder of this note is organized as follows. In Section~\ref{S:phase} we map out the phase diagram of both sides of~\eqref{E:proposal}, and show that not only can we match the TFTs describing the massive phases, but we can also match the Chern-Simons terms for the global symmetries. We match baryons, monopoles, and global symmetries in Section~\ref{S:global}. Gauging an appropriate $U(1)$ subgroup of the global symmetry on both sides of our proposal~\eqref{E:proposal}, we find that~\eqref{E:proposal} also implies a $U/U$ duality which we describe in Section~\ref{S:Uduality}. In Section~\ref{S:real} we comment briefly on an extension of our proposal~\eqref{E:proposal} to dualities between Chern-Simons theories with $SO$ and $USp$ gauge groups, and we conclude with some open questions in Section~\ref{S:conclude}.

\emph{Note:} While this work was nearing completion F.~Benini posted a paper~\cite{Benini:2017aed} which also conjectures the duality~\eqref{E:proposal} as well as extensions for other classical gauge groups.

\section{Mapping out the phase diagram}
\label{S:phase}

We begin with the Lagrangians for the theories on both sides of our proposed duality~\eqref{E:proposal}. We work in Euclidean signature. The $SU(N)_{-k+\frac{N_f}{2}}$ theory is described by a Lagrangian\footnote{In this work we follow~\cite{Hsin:2016blu} and use the convention that the functional determinant of the Dirac operator of a single Dirac fermion coupled to an external gauge field $A$ and metric $g$ is given by
\beq
	\text{det}(\slashed{D}(A,g)) = |\text{det}(\slashed{D}(A,g))| \exp\left( -\frac{i\pi\eta(A,g)}{2}\right)\,,
\eeq
where $\eta$ is the $\eta$-invariant. On a closed manifold with trivial topology, this phase evaluates to a Chern-Simons term with level $\frac{1}{2}$ for $A$ along with a gravitational Chern-Simons term,
\beq
	\frac{i \pi \eta(A,g)}{2} \to i \int\left\{\frac{1}{8\pi}AdA+\frac{1}{196\pi}\text{tr}\left( \Gamma d\Gamma+\frac{2}{3}\Gamma^3\right)\right\}\,.
\eeq
Giving the fermion a mass and integrating it out, one finds the usual one-loop exact shifts to the Chern-Simons level in the infrared. With this convention, a positive Dirac mass does not shift the bare level, while a negative mass shifts the level by $-1$. When $A$ is dynamical with a bare Chern-Simons level $k$, we refer to the massless theory as $U(1)_{k-\frac{1}{2}}$. }
\beq
	\mathcal{L}_{SU} =-i\frac{-k+N_f}{4\pi} \text{tr}\left( ada-\frac{2i}{3}a^3\right) + \bar{\psi}^i i \slashed{D} \psi_i + (D^{\mu}\phi^{\dagger m})(D_{\mu}\phi_m) + \mathcal{L}_{\rm int}\,,
\eeq
where $a$ is the $SU(N)$ gauge field, $\mathcal{L}_{\rm int}$ describes scalar and fermion interactions, $i=1,..,N_f$ is a fermion flavor index, and $m=1,..,N_s$ is a scalar flavor index. This theory has a manifest $U(N_f)\times U(N_s)$ global symmetry which we impose. To get a handle on the scalar and fermion interactions we consider two limits. The first is to realize the $SU$ theory as an infrared fixed point of a renormalization group flow, starting with a free theory in the UV. For general $N_f$, $N_s$, the classically relevant and marginal operators are
\beq
	\label{E:potential}
	|\phi|^2\,, \qquad \bar{\psi}\psi\,, \qquad |\phi|^4\,, \qquad |\phi^6| \,, \qquad (\bar{\psi} \psi)|\phi|^2\,, \qquad (\bar{\psi}^i\cdot \phi_m)(\phi^{\dagger m}\cdot \psi_i)\,.
\eeq
These operators must be tuned to reach a non-trivial IR CFT. In the IR the scalar $\phi$ becomes a ``Wilson-Fisher'' scalar, and one expects $|\phi|^4$ and $|\phi^6|$ to both be irrelevant with respect to the IR scalings. 

What of the quartic fermion/scalar operators? Now consider a large $N$ limit. At large $N$, the fermion and Wilson-Fisher scalar both have dimension $1$. The operators $|\phi|^4$, $(\bar{\psi}\cdot \psi)|\phi|^2$, and $|\phi|^6$ are then ``multi-trace'' operators whose dimensions by large-$N$ factorization are $4+O(N^{-1})$ and $6+O(N^{-1})$. However it is easy to check that the last quartic operator,
\beq
\mathcal{O}_4 = (\bar{\psi}^i \cdot \phi_m)(\phi^{\dagger m}\cdot \psi_i)\,,
\eeq
remains approximately marginal with $\Delta = 3 + O(N^{-1})$. This operator will play a crucial role in what follows. 

Now consider the $U(k)_{N-\frac{N_s}{2}}$ theory. Its Lagrangian is
\beq
	\mathcal{L}_U =-i \frac{N}{4\pi}\text{tr}\left( a'da'-\frac{2i}{3}a'^3\right) + (D^{\mu}\Phi^{\dagger i})(D_{\mu} \Phi_i) + \bar{\Psi}^m i \slashed{D} \Psi_m + \mathcal{L}_{\rm int}'\,,
\eeq
where $a'$ is the $U(k)$ gauge field, and again $i=1,..,N_f$, $m=1,..,N_s$ and $\mathcal{L}_{\rm int}'$ describes the scalar and fermion interactions. We denote the scalars of the $U$ theory as $\Phi_i$ and the fermions as $\Psi_m$ to distinguish them from the bosons and fermions of the $SU$ theory. This theory has a manifest $SU(N_s)\times SU(N_f)\times U(1)_m \times U(1)$ global symmetry. The $U(1)_m$ is a monopole number, while $U(1)$ is carried by the fundamental fermions (with charge $+1$) and bosons (with charge $-1$). As in the $SU$ theory, we expect that the operators in $\mathcal{L}_{\rm int}'$ are 
\beq
|\Phi|^2\,, \qquad \bar{\Psi}\Psi\,, \qquad |\Phi|^4\,, \qquad |\Phi|^6\,, \qquad (\bar{\Psi} \Psi)|\Phi|^2\,, \qquad (\bar{\Psi}^m \cdot \Phi_i)(\Phi^{\dagger i}\cdot \Psi_m)\,,
\eeq
whose coefficients are all tuned to realize a non-trivial IR CFT. As above, the operator
\beq
\mathcal{O}_4' = (\bar{\Psi}^m \cdot \Phi_i)(\Phi^{\dagger i}\cdot \Psi_m)\,,
\eeq
will soon reveal its importance.

\subsection{Massive phases and critical lines}

At large $N$ the only relevant $SU(N_f)\times SU(N_s)\times U(1)\times U(1)$-invariant operators in the $SU$ and $U$ theories are the scalar and fermion mass operators. To simplify our analysis we subsequently assume that this remains true for finite $N$, although this assumption may be wrong. The quartic operator $\mathcal{O}_4$ has dimension $3+O(1/N)$ at large $N$ and so may be relevant, depending on the sign of the $1/N$ correction. We also assume that all phases are the ones accessible semiclassically at large $|m_{\psi}|, |m_{\phi}^2|$ when realizing these theories as the endpoint of flows starting from a UV free-field fixed point. Finally, for the purposes of a simple presentation, we assume that there are no first order transitions, with the caveat that there could very well be first order transitions separating the semiclassical phases we find at large $|m_{\psi}|$, $|m_{\phi}|^2$ from the region at small mass. Subject to these assumptions we map out the schematic two-dimensional phase diagram of both theories.

Let us begin with the $SU$ theory. We can give the fermions a positive or negative mass, as well as give the scalars a positive or negative mass-squared. Integrating out the massive fermions leads to a one-loop exact shift of the Chern-Simons level. Giving a positive mass-squared simply decouples the scalars, while a negative mass-squared partially Higgses the gauge symmetry. Following previous work, we assume that the interactions prefer to maximally Higgs the gauge theory from $SU(N)$ down to $SU(N-N_s)$, and find that this is necessary in order for our conjecture to work. 

In the Higgsed phase, the quartic operators
\beq
(\bar{\psi}\psi)|\phi|^2\,, \qquad \mathcal{O}_4 = (\bar{\psi}^i \cdot \phi_m)(\phi^{\dagger m}\cdot \psi_i)\,,
\eeq
effectively generate fermion masses. The first gives a mass to all fermions, which can be compensated for by a suitable shift of the coupling of the fermion mass operator $\bar{\psi}\psi$. The second is more interesting. The original $N N_f$ fermions break up into $N_f$ fundamental representations of the unbroken $SU(N-N_s)$ gauge symmetry, while the remaining $N_s N_f$ fermions are gauge-singlets. We refer to these as singlet fermions. Crucially, the operator $\mathcal{O}_4$ generates a mass $|\phi|^2$ for the singlet fermions in the Higgs phase.

Denote the coefficient multiplying $\mathcal{O}_4$ in the Higgs phase as $c_4$. If $c_4$ is positive, then this operator contributes a positive mass to the singlet fermions, while if $c_4$ is negative it contributes a negative mass. In either case, since the fermion mass operator $\bar{\psi}\psi$ gives a mass to all fermions, it is clear that there are \emph{three} distinct massive Higgs phases. In one all fermions have a positive mass, in another all fermions have a negative mass, and in the last the singlet fermions have a mass whose sign is opposite those of the remaining charged fermions. We will soon see that our duality requires $c_4<0$, so that this new phase exists for $m_{\psi}>0$ and that in this phase the singlet fermions have a negative mass. 

In what follows we must also distinguish between the cases $N_s < N$ and $N_s = N$. For $N_s < N$ and away from critical lines the Higgsed phase of the $SU$ theory is completely gapped. However for $N_s = N$ the Higgsed phase is gapless with a massless scalar. Correspondingly, in the $U$ theory, for $N_s<N$ both signs of the Dirac mass lead to a non-trivial TFT in the infrared. For $N_s = N$ a negative Dirac mass leaves a $U(k)_0$ or $U(k-N_f)_0$ theory (depending on whether or not one is in the Higgsed phase), whose non-abelian part confines at low energies leaving behind compact electromagnetism in the IR, not a TFT. Ultimately, we will find that our proposal is still consistent for $N_s = N$, however only when $k <N_f$ rather than $k\leq N_f$. That is we must also have the flavor bound $(N_f,N_s)\neq (k,N)$ as advertised in the Introduction.

\subsubsection{$N_s<N$}

\begin{figure}[t]
\centering
\subfloat[]{\includegraphics[width=3in]{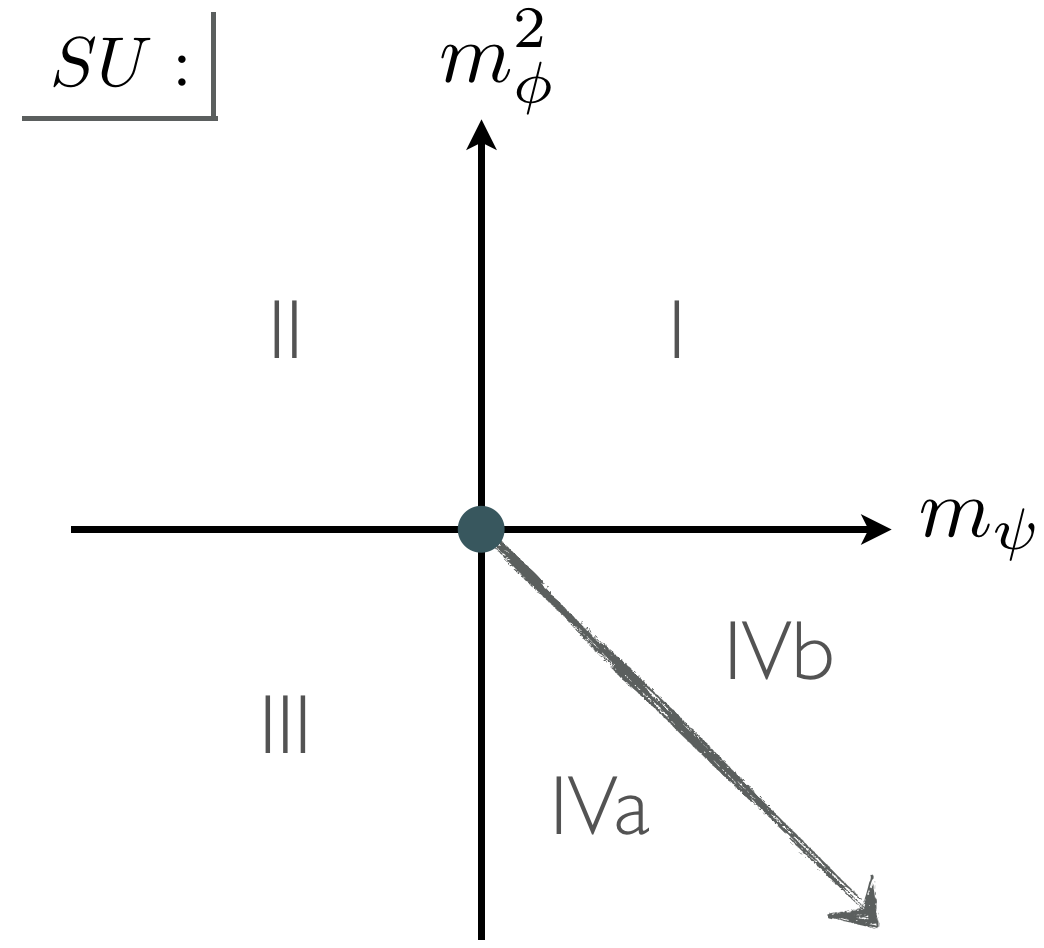}}
\hfill
\subfloat[]{\includegraphics[width=3.1in]{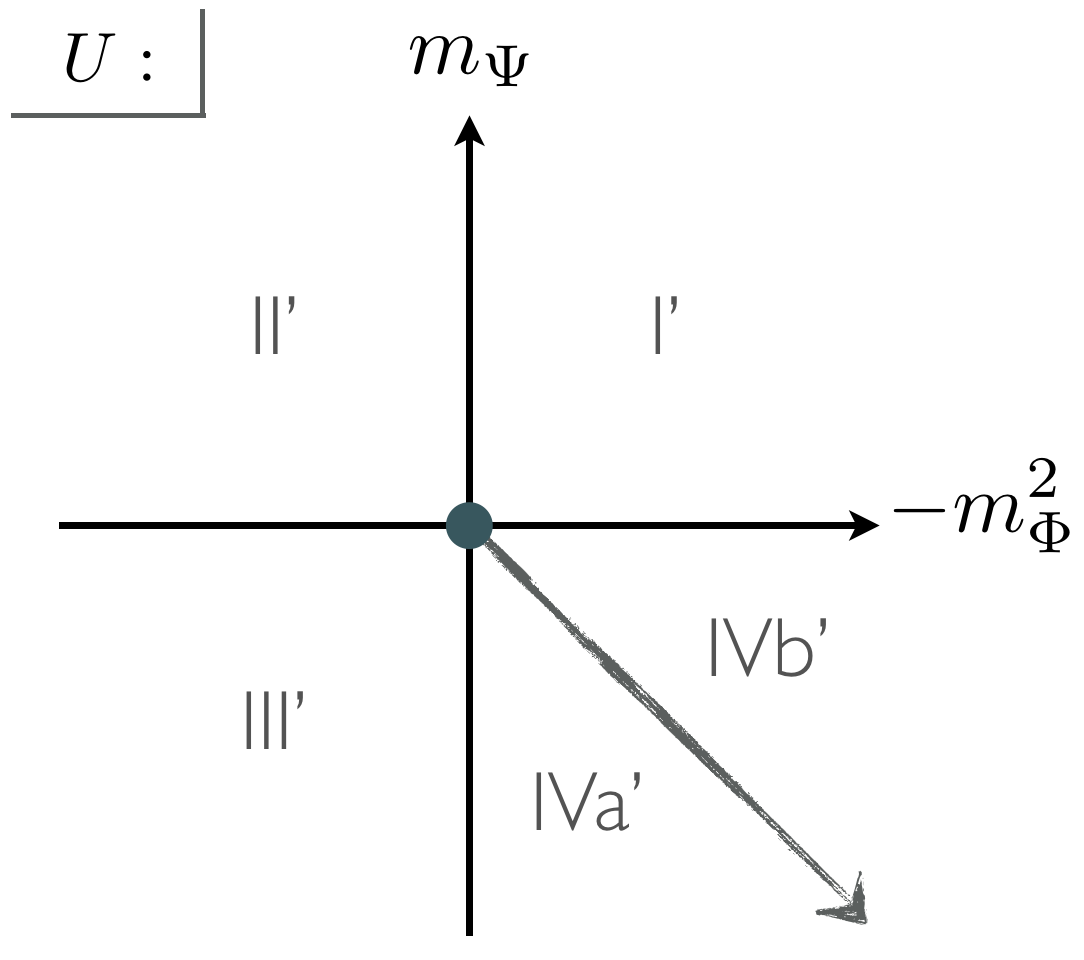}}
\caption{\label{F:phases} For $N_s <N$, the schematic phase diagrams of (a) $SU(N)_{-k+\frac{N_f}{2}}$ Chern-Simons theory coupled to $N_f$ fermions and $N_s$ scalars and (b) $U(k)_{N-\frac{N_s}{2}}$ Chern-Simons theory coupled to $N_s$ fermions and $N_f$ scalars. The various phases are described by IR TFTs given in~\eqref{E:SUphases} and~\eqref{E:Uphases}. and the critical lines are described by the CFTs given in~\eqref{E:SUlines} and~\eqref{E:Ulines}.}
\end{figure}

We begin with the case $N_s < N$. There are \emph{five} distinct massive phases separated by various critical lines, with the critical $SU$ theory living at the original of the phase diagram. However there are only four different IR TFTs. See Figure~\ref{F:phases}. The distinct TFTs governing the massive phases are
\begin{align}
\begin{split}
	\label{E:SUphases}
	(\text{I}) \quad m_{\psi}>0\,, \, m_{\phi}^2>0 \,: &\quad SU(N)_{-k+N_f} \,,
	\\
	(\text{II}) \quad m_{\psi}<0\,, \, m_{\phi}^2>0 \,: &\quad SU(N)_{-k}\,,
	\\
	(\text{III}) \quad m_{\psi}<0\,, \, m_{\phi}^2<0 \,: &\quad SU(N-N_s)_{-k}\,,
	\\
	(\text{IV}) \quad m_{\psi}>0\,, \, m_{\phi}^2<0\, : & \quad SU(N-N_s)_{-k+N_f}\,.
\end{split}
\end{align}
Observe that $N_s <N$ is required to have a massive Higgs phase. For $k=N_f$ Phases I and IV are described by $SU(N)_0$ and $SU(N-N_s)_0$, which give confining Yang-Mills theories rather than TFTs. The fourth phase splits into two, one at large positive $m_{\psi}$ where all fermions have a positive mass, and one at intermediate positive $m_{\psi}$, where the singlet fermions have a negative mass,
\begin{align}
\begin{split}
	\label{E:SUphases4}
	(\text{IVa}) \quad m_{\psi}>0\,, \, m_s <0 \,, \, m_{\phi}^2<0 \,: & \quad SU(N-N_s)_{-k+N_f}\,,
	\\
	(\text{IVb}) \quad m_{\psi}>0\,, \, m_s>0 \,, \, m_{\phi}^2 <0 \, : & \quad SU(N-N_s)_{-k+N_f}\,.
\end{split}
\end{align}
Now for the critical lines. There are five critical lines separating the various phases. Four of the lines are described by $SU$ Chern-Simons theory coupled to either fermions or bosons, and one is the theory of massless singlet fermions. We summarize the theories on the critical lines as
\begin{align}
\begin{split}
	\label{E:SUlines}
	(\text{I-II}) \quad &SU(N)_{-k+\frac{N_f}{2}} \text{ with }N_f \, \psi\,,
	\\
	(\text{II-III}) \quad &SU(N)_{-k} \text{ with }N_s\, \phi\,,
	\\
	(\text{III-IVa}) \quad &SU(N-N_s)_{-k+\frac{N_f}{2}} \text{ with }N_f\, \psi\,,
	\\
	(\text{IVa-IVb}) \quad & N_s N_f \text{ singlet } \psi\text{'s} \, + \, SU(N-N_s)_{-k+N_f} \text{ TFT}\,,
	\\
	(\text{IVb-I}) \quad & SU(N)_{-k+N_f} \text{ with }N_s \,\phi\,.
\end{split}
\end{align}
For the case $k=N_f$, we see that Phases I and IV are trivial; the line between Phases IVa and IVb only contains the singlet fermions; and the line between phase IVb and I is $SU(N)$ Yang-Mills coupled to $N_s$ scalars, which we expect to confine and lead to a gapped theory.

By the by, the operator $\mathcal{O}_4$ must be present in order that the singlet fermions are gapped on then critical line separating Phases III and IVa. However, this requirement does not constrain the sign of its coupling $c_4$. We do note that if $c_4$ were positive, then the critical line with the massless singlet fermions would run through Phase III rather than Phase IV.

Having mapped out the $SU$ phase diagram we move on to consider the $U$ theory. Much of the discussion of the $U$ phase diagram carries over here without modification, and so let us simply summarize the salient features. See Figure~\ref{F:phases}. As before, there are five distinct massive phases described by four different IR TFTs, which are given by
\begin{align}
\begin{split}
	\label{E:Uphases}
	(\text{I'}) \quad m_{\Psi}>0\,, \, m_{\Phi}^2<0 \,: &\quad U(k-N_f)_{N} \,,
	\\
	(\text{II'}) \quad m_{\Psi}>0\,, \, m_{\Phi}^2>0 \,: &\quad U(k)_{N}\,,
	\\
	(\text{III'}) \quad m_{\Psi}<0\,, \, m_{\Phi}^2>0 \,: &\quad U(k)_{N-N_s}\,,
	\\
	(\text{IV'}) \quad m_{\Psi}<0\,, \, m_{\Phi}^2<0\, : & \quad U(k-N_f)_{N-N_s}\,.
\end{split}
\end{align}
To have a massive Higgs phase we must respect $N_f \leq k$.  In a moment we will see that the coefficient of $\mathcal{O}_4'$, $c_4'$, must be positive, so that Phase IV splits into two, with
\begin{align}
\begin{split}
	\label{E:Uphases4}
	(\text{IVa'}) \quad m_{\Psi}<0\,, \, m_s <0 \,, \, m_{\Phi}^2<0 \,: & \quad U(k-N_f)_{N-N_s}\,,
	\\
	(\text{IVb'}) \quad m_{\Psi}<0\,, \, m_s>0 \,, \, m_{\Phi}^2 <0 \, : & \quad U(k-N_f)_{N-N_s}\,.
\end{split}
\end{align}
The critical $U$ theory sits at the origin of the phase diagram, and the critical lines separating the massive phases are described by
\begin{align}
\begin{split}
	\label{E:Ulines}
	(\text{I'-II'}) \quad & U(k)_{N} \text{ with }N_f \, \Phi\,,
	\\
	(\text{II'-III'}) \quad & U(k)_{N-\frac{N_s}{2}} \text{ with }N_s\, \Psi\,,
	\\
	(\text{III'-IVa'}) \quad & U(k)_{N-N_s} \text{ with }N_f\, \Phi\,,
	\\
	(\text{IVa'-IVb'}) \quad & N_s N_f \text{ singlet } \Psi\text{'s} \, + \, U(k-N_f)_{N-N_s}\text{ TFT}\,,
	\\
	(\text{IVb'-I'}) \quad & U(k-N_f)_{N-\frac{N_s}{2}} \text{ with }N_s\, \Psi\,.
\end{split}
\end{align}
If $c_4'$ were negative, the critical line with the massless singlet fermions would run through Phase I' rather than Phase IV'. For the special case $k=N_f$, we see that Phases I' and IV' are trivial, that the line separating Phases IVa' and IVb' is just given by the massless singlet fermions, and the line between Phase IVb' and I' is trivial.

Having assembled all of this information, we may match the phases, lines, and operators of the two theories. Using the basic level/rank duality~\eqref{E:levelrank} equating
\begin{equation*}
	SU(N)_{-k} \quad \leftrightarrow \quad U(k)_N\,,
\end{equation*}
we see that the TFTs governing the massive phases of the $SU$ theory~\eqref{E:SUphases} are precisely those describing the massive phases of the $U$ theory~\eqref{E:Uphases}. Phase I maps to Phase I', and similarly for the others. Comparing the axes on the two phase diagrams in Figure~\ref{F:phases} then tells us how the mass operators map under the duality, with
\beq
	\bar{\psi} \psi \leftrightarrow - |\Phi|^2\,, 
	\qquad
	|\phi|^2 \leftrightarrow \bar{\Psi} \Psi\,.
\eeq
Furthermore, we see that the critical lines in both theories~\eqref{E:SUlines} and~\eqref{E:Ulines} match upon using Aharony's dualities~\eqref{E:sequence1} and~\eqref{E:sequence2}. To match the lines running through Phases IV and IV' we also require level/rank duality to equate the TFTs arising on each line. 

For the special case $k=N_f$, the massive phases still match on account of the fact that Phases I, I', IV, and IV' are all trivial. The critical lines also match, upon recalling that the IVb-I and IVb'-I' lines are trivial.

We also see that the assumption that our duality holds determines the sign of the quartic couplings $c_4$ and $c_4'$. As we mentioned above, if $c_4$ were positive, the ``singlet critical line'' of the $SU$ theory would run through Phase III, and if $c_4'$ were negative, the singlet critical line of the $U$ theory would run through Phase I'. The only consistent possibility is $c_4<0, c_4'>0$. An even simpler argument is that Phase IV is the only quadrant of the phase diagram in which both the $SU(N)$ theory and its $U(k)$ dual are both in a Higgs phase, and so the singlet line must run through it. As we mentioned in the Introduction, we see that the quartic coupling has the same sign as the Chern-Simons level.

\subsubsection{$N_s=N$}

\begin{figure}[t]
\centering
\subfloat[]{\includegraphics[width=3in]{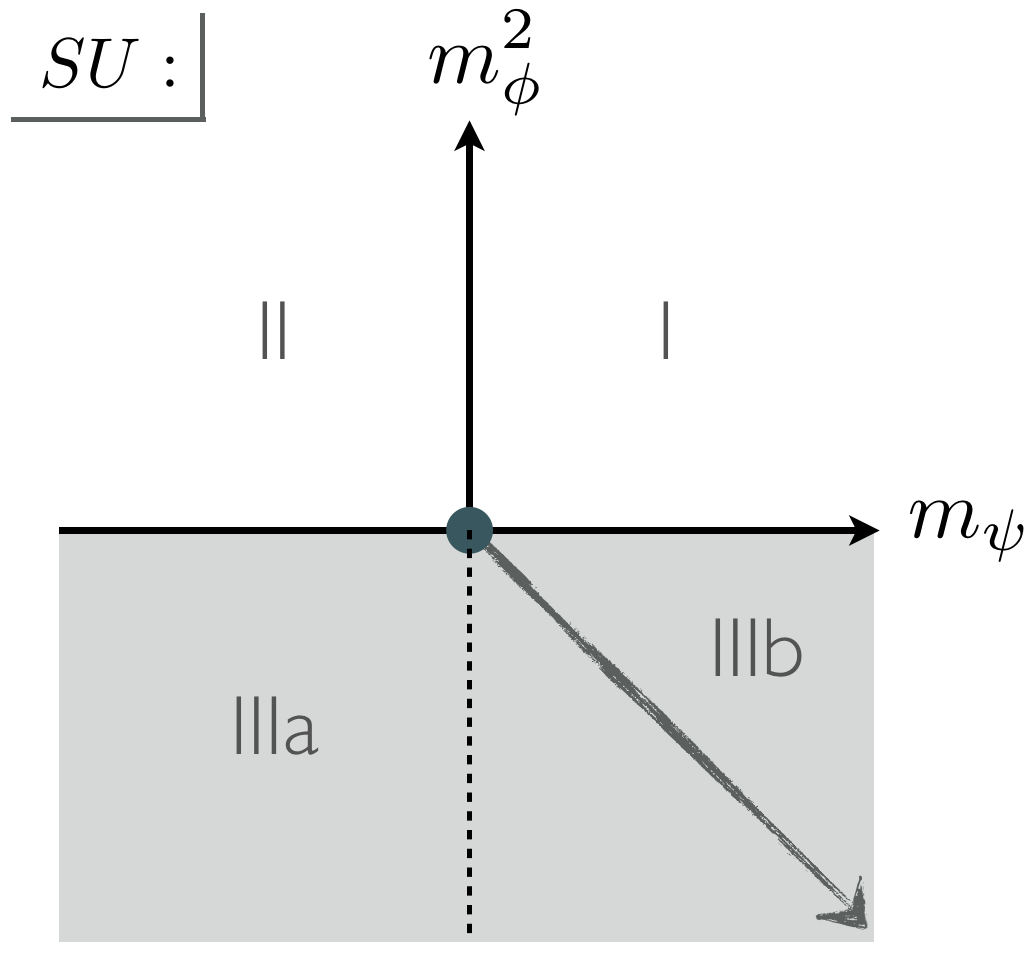}}
\hfill
\subfloat[]{\includegraphics[width=3.1in]{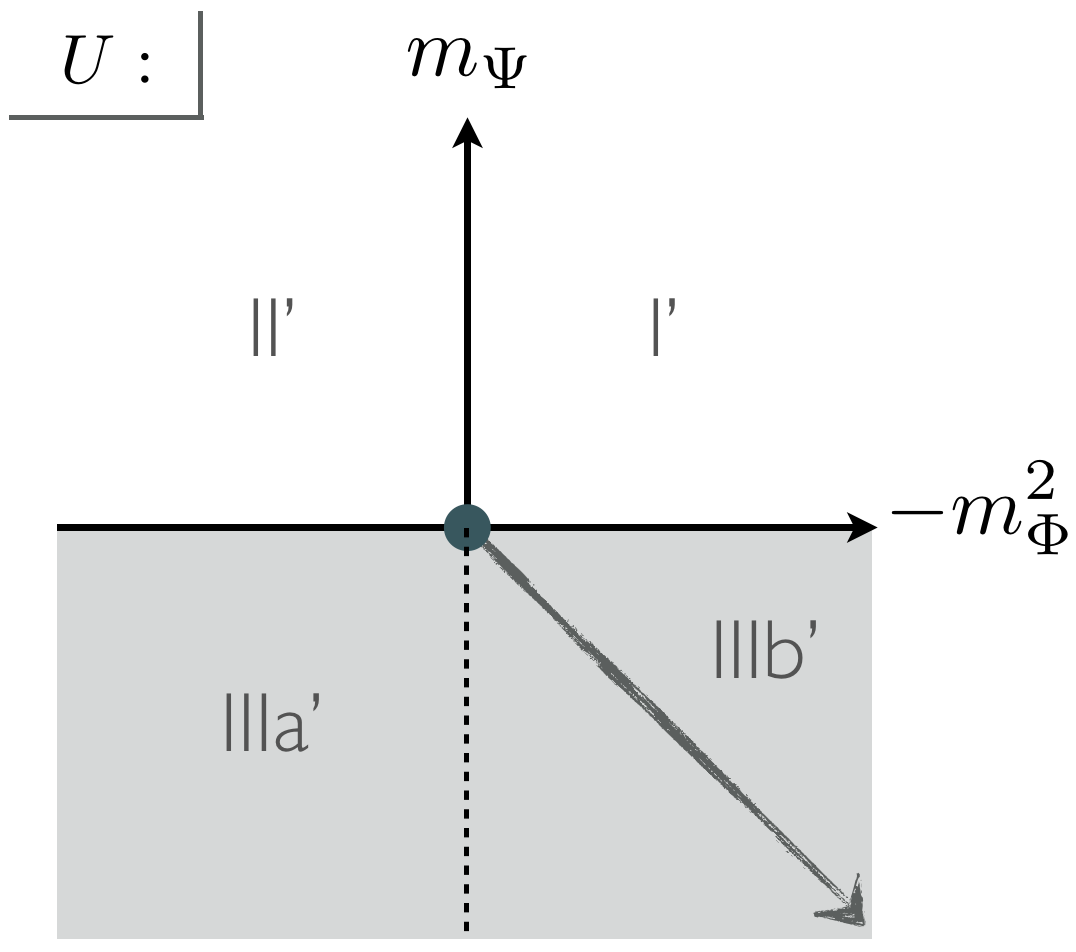}}
\caption{\label{F:phases2} For $N_s =N$ and $N_f<k$, the schematic phase diagrams of (a) $SU(N)_{-k+\frac{N_f}{2}}$ Chern-Simons theory coupled to $N_f$ fermions and $N_s$ scalars and (b) $U(k)_{N-\frac{N_s}{2}}$ Chern-Simons theory coupled to $N_s$ fermions and $N_f$ scalars. The various phases are described by~\eqref{E:SUphases2} and~\eqref{E:Uphases2}. and the critical lines are described by the CFTs given in~\eqref{E:SUlines2} and~\eqref{E:Ulines2}. The shaded region is gapless.}
\end{figure}

Now we tackle the case $N_s = N$. We will be brief and summarize the main features. See Figures~\ref{F:phases2} for the schematic phase diagrams when $N_f<k$. 

In the $SU$ theories there are four phases (not all of which are massive). In the Higgs phase we have $SU(N) \to 0$, and the only remnant of the scalars is a single compact Goldstone boson. Dualizing it into pure electromagnetism, we write the various phases as
\begin{align}
\begin{split}
\label{E:SUphases2}
	(\text{I}) \quad m_{\psi}>0\,, \, m_{\phi}^2>0 \,: &\quad SU(N)_{-k+N_f} \,,
	\\
	(\text{II}) \quad m_{\psi}<0\,, \, m_{\phi}^2>0 \,: &\quad SU(N)_{-k}\,,
	\\
	(\text{III}) \hspace{.75in} m_{\phi}^2<0 \,: &\quad U(1)_0\,,
\end{split}
\end{align}
and Phase III splits into Phases IIIa and IIIb depending on whether the fermions (note that since the gauge group is trivial in the Higgs phase \emph{all} fermions are ``singlets.'') get a negative or positive mass. For $k=N_f$, the theory in Phase I is $SU(N)_0$ which we expect to be massive. The critical lines separating the phases are
\begin{align}
\begin{split}
	\label{E:SUlines2}
	(\text{I-II}) \quad &SU(N)_{-k+\frac{N_f}{2}} \text{ with }N_f \, \psi\,,
	\\
	(\text{II-III}) \quad &SU(N)_{-k} \text{ with }N\, \phi\,,
	\\
	(\text{IIIa-IIIb}) \quad & NN_f \text{ free } \psi\text{'s}  \, + \, \text{ decoupled }U(1)_0\,,
	\\
	(\text{IIIb-I}) \quad & SU(N)_{-k+N_f} \text{ with }N \,\phi\,.
\end{split}
\end{align}
For $k=N_f$ Phase I and the line separating Phases IIIb and I are trivial.

For $N_f<k$ the $U$ theories also have four phases, while for $N_f = k$ they have five. We have
\begin{align}
\begin{split}
	\label{E:Uphases2}
	(\text{I'}) \quad m_{\Psi}>0\,, \, m_{\Phi}^2<0 \,: &\quad U(k-N_f)_{N} \,,
	\\
	(\text{II'}) \quad m_{\Psi}>0\,, \, m_{\Phi}^2>0 \,: &\quad U(k)_{N}\,,
	\\
	(\text{III'}) \quad m_{\Psi}<0\,, \, m_{\Phi}^2>0 \,: &\quad U(k)_0 \to U(1)_0\,,
	\\
	(\text{IV'}) \quad m_{\Psi}<0\,, \, m_{\Phi}^2<0\, : & \quad U(k-N_f)_0 \to \begin{cases} U(1)_0\,, & N_f<k \\ 0 \,, & k=N_f\,,\end{cases}
\end{split}
\end{align}
and Phase IV' only exists when $N_f = k$. For $N_f < k$, Phase III' splits into Phases IIIa' and IIIb' depending on whether the fermions have a negative or positive mass. For $N_f = k$, Phase IV' splits into Phases IVa' and IVb' with the same. For Phases III' and, when it exists, IV', we are using the expectation that the non-abelian part of $U(k)_0$ confines at low energies so that its low energy limit is described by pure electromagnetism. The critical lines when $N_f <k$ are
\begin{align}
\begin{split}
	\label{E:Ulines2}
	(\text{I'-II'}) \quad & U(k)_{N} \text{ with }N_f \, \Phi\,,
	\\
	(\text{II'-IIIa'}) \quad & U(k)_{N-\frac{N}{2}} \text{ with }N\, \Psi\,,
	\\
	(\text{IIIa'-IIIb'}) \quad  & N N_f\text{ free } \Psi\text{'s} \, +  \text{ decoupled } U(1)_0\,, 
	\\
	(\text{IIIb'-I'}) \quad & U(k-N_f)_{N-\frac{N}{2}} \text{ with }N\, \Psi\,.
\end{split}
\end{align}

Comparing the phases of the $SU$ theory~\eqref{E:SUphases2} and the phases of the $U$ theory~\eqref{E:Uphases2}, we see that the two perfectly match provided that $N_f<k$. The same is true for the critical lines. However, clearly there is no match when $k=N_f$, and thus we demand the last of our flavor bounds, $(N_f,N_s)\neq (k,N)$.

\subsection{Matching flavor Chern-Simons terms}

In the last Subsection we matched the various massive phases and critical lines of the $SU$ and $U$ theories appearing in our proposed duality~\eqref{E:proposal}. In fact we can perform an even stronger test. We first couple both sides to slowly varying background gauge fields. (We could also put the theories on a spin manifold with a slowly varying metric, but we do not do so in this work.) The massive phases are then described not only by non-trivial TFTs, but there are also flavor Chern-Simons terms whose levels are one-loop exact and which we may match. It is straightforward to deduce the levels for the non-abelian part of the flavor symmetry, but as we will see, we must be careful when computing the levels for the abelian part. It is particularly tricky to compute the abelian levels in the $SU$ theories, and we will find it convenient to represent the $SU(N)$ theories as $U(N)$ theories subject to a $U(1)$ quotient.

\subsubsection{The $U(k)$ theories}

We begin with the $U(k)$ theories. The manifest Abelian global symmetry of these theories is $U(1)_m \times U(1)_F$, where $U(1)_m$ refers to a monopole number and the $U(1)_F$ is an ordinary charge under which the bosons carry charge $0$ and the fermions charge $-1$. With this convention, the manifest global symmetry is in fact $U(N_f)\times SU(N_s)\times U(1)_m$. Turning on background gauge fields which couple the manifest symmetry and including carefully chosen Chern-Simons terms for the external fields, the Lagrangian is
\beq
	\label{E:Ulagrangian}
	\mathcal{L}_U = -i \left[ \frac{N}{4\pi}\text{tr}\left( a' da'-\frac{2i}{3}a'^3\right)+ \frac{1}{2\pi}\text{tr}(a') d\tilde{A}_1' \right]+ |D\Phi|^2 + \bar{\Psi} i \slashed{D} \Psi + \mathcal{L}_{\rm int}' \,.
\eeq
together with
\begin{align}
\begin{split}
	D_{\mu}\Phi & = \left( \partial_{\mu} - i ( a'_{\mu} \mathbbm{1}_f + A'_{\mu} \mathbbm{1}_c)\right)\Phi\,,
	\\
	D_{\mu}\Psi & = \left( \partial_{\mu} - i ( a'_{\mu} \mathbbm{1}_f + B'_{\mu} \mathbbm{1}_c- \tilde{A}_{2\mu}'\mathbbm{1} )\right)\Psi\,.
\end{split}
\end{align}
Here $a'_{\mu}$ is a $U(k)$ gauge field, $A'_{\mu}$ is a background $SU(N_f)$ gauge field, and $B'_{\mu}$ a background $SU(N_s)$ gauge field. There are also background abelian gauge fields: the $U(1)_m$ gauge field is $\tilde{A}_{1\mu}'$ and the $U(1)_F$ gauge field is $\tilde{A}_{2\mu}'$. (We can group $B'$ and $\tilde{A}_2'$ into a $U(N_s)$ gauge field if we wish.) Observe that $\tilde{A}_1'$ only appears through a $BF$ coupling to the monopole current $\star j_m = \frac{1}{2\pi}d\text{tr}(a')$.

The various abelian Chern-Simons levels are subject to quantization conditions, which when violated characterize 't Hooft anomalies for the flavor symemtries. For now, we will simply proceed to compute the levels in the various phases without worrying about the precise quantization conditions. Furthermore, we are being a bit sloppy in writing~\eqref{E:Ulagrangian}. For generic parameters we are not allowed to set the various abelian levels to vanish. What we are really doing in this Subsection is to compute the jumps in flavor Chern-Simons levels from one phase to another, and these jumps of course do not depend on these details.

The five massive phases~\eqref{E:Uphases} are obtained after turning on fermion and scalar masses. The various Chern-Simons levels receive one-loop shifts after integrating out fermions, as well as additional shifts in the Higgsed phases. Including the flavor groups, the massive phases are characterized by
\begin{align}
\begin{split}
	\label{E:Uflavor}
	(\text{I'}) \, : \quad & U(k-N_f)_N \times SU(N_f)_{N} \times SU(N_s)_{0} \times J_{\text{I'}}\,,
	\\
	(\text{II'}) \, : \quad & U(k)_N \times SU(N_f)_{0} \times SU(N_s)_{0} \times J_{\text{II'}} \,,
	\\
	(\text{III'}) \, : \quad & U(k)_{N-N_s} \times SU(N_f)_{0} \times SU(N_s)_{-k} \times J_{\text{III'}} \,,
	\\
	(\text{IVa'}) \, : \quad & U(k-N_f)_{N-N_s} \times SU(N_f)_{N-N_s} \times SU(N_s)_{-k} \times J_{\text{IVa'}} \,,
	\\
	(\text{IVb'}) \, : \quad & U(k-N_f)_{N-N_s} \times SU(N_f)_{N} \times SU(N_s)_{-k+N_f} \times J_{\text{IVb'}}\,,
\end{split}
\end{align}
where only the first group is dynamical, and there are $2\times 2$ matrices describing the abelian Chern-Simons levels in each phase,
\beq
\frac{J'^{ab}}{4\pi}\tilde{A}_a' d\tilde{A}_b'\,.
\eeq

We need to integrate out the dynamical $U(1)$ gauge field $\text{tr}(a')$ to get these abelian Chern-Simons levels. For example, in Phase I', $\text{tr}(a')$ appears through two terms in the low-energy effective Lagrangian,
\beq
	\label{E:Uphase1A}
	\mathcal{L}_{\text{tr}(a')} = -i \left[ \frac{ N}{4\pi (k-N_f)} \text{tr}(a')d\text{tr}(a') + \frac{1}{2\pi}\text{tr}(a')d\tilde{A}_1'\right]\,.
\eeq
(The first term is the abelian part of the $U(k-N_f)$ Chern-Simons term.) There is effectively an equation of motion for $\text{tr}(a')$ which sets
\beq
	\text{tr}(a') = \frac{-k+N_f}{N} \tilde{A}_1'\,,
\eeq
so that~\eqref{E:Uphase1A} becomes
\beq
	\mathcal{L}_{\text{tr}(a')} \to - i \frac{-k + N_f}{4\pi N}\tilde{A}'_1 d\tilde{A}'_1\,,
\eeq 
i.e. an effective Chern-Simons term for $\tilde{A}_1'$ at level $ \frac{- k+N_f}{N}$. Accounting for the one-loop shifts to the bare levels (which happen to vanish in this phase), the $2\times 2$ matrix of abelian CS levels is
\begin{subequations}
\beq
	\label{E:Uabelian1}
	J_{\text{I'}}^{ab} = \begin{pmatrix} \frac{-k+N_f}{N} & 0 \\ 0 & 0  \end{pmatrix}\,.
\eeq
Similar computations in the other phases of the $U$ theory give
\begin{align}
\begin{split}
	\label{E:Uabelian2}
	J_{\text{II'}}^{ab} & = \begin{pmatrix} -\frac{k}{N} & 0 \\ 0 & 0 \end{pmatrix}\,,
	\\
	J_{\text{III'}}^{ab} & = -\frac{k}{N-N_s}\begin{pmatrix} 1 & N_s \\ N_s & N N_s\end{pmatrix}\,,
	\\
	J_{\text{IVa'}}^{ab} & = -\frac{k-N_f}{N-N_s}\begin{pmatrix} 1&  N_s \\ N_s & NN_s \end{pmatrix}-N_f N_s \begin{pmatrix}0 & 0 \\ 0 & 1 \end{pmatrix}\,,
	\\
	J_{\text{IVb'}}^{ab} & =  -\frac{k-N_f}{N-N_s}\begin{pmatrix} 1&  N_s \\ N_s & NN_s \end{pmatrix}\,.
\end{split}
\end{align}
\end{subequations}

\subsubsection{The $SU(N)$ theories}

We continue with the $SU(N)$ theories. Turning on a background for the manifest $U(N_f)\times U(N_s) = \big(SU(N_f)\times U(1)_f\big)/\mathbb{Z}_{N_f}\times \big(SU(N_s)\times U(1)_s\big)/\mathbb{Z}_{N_s}$ global symmetry, and including Chern-Simons terms for the background fields, the Lagrangian is
\begin{align}
\begin{split}
	\label{E:SUlagrangian}
	\mathcal{L}_{SU} &= -i \frac{-k+N_f}{4\pi}\text{tr}\left(ada-\frac{2i}{3}a^3\right) + \bar{\psi}i\slashed{D}\psi + |D\phi|^2 + \mathcal{L}_{\rm int}
	\\
	& \qquad - i \left[ \frac{N}{4\pi}\text{tr}\left( A dA-\frac{2i}{3}A^3\right)  + \frac{J^{ab}}{4\pi}\tilde{A}_ad\tilde{A}_b\right]\,,
\end{split}
\end{align}
along with
\begin{align}
\begin{split}
	D_{\mu}\psi &= \left( \partial_{\mu} -i \left(a_{\mu} \mathbbm{1}_f  + A_{\mu}\mathbbm{1}_c +\frac{1}{N} \tilde{A}_{1\mu} \mathbbm{1}\right)\right)\psi\,,
	\\
	D_{\mu}\phi & = \left(\partial_{\mu} - i \left( a_{\mu}\mathbbm{1}_f + B_{\mu}\mathbbm{1}_c + \left(\frac{1}{N}\tilde{A}_{1\mu}+\tilde{A}_{2\mu}\right)\mathbbm{1}\right)\right)\phi\,,
\end{split}
\end{align}
where $\mathbbm{1}_f$ acts as the identity on flavor indices, $\mathbbm{1}_c$ as the identity on color indices, and $\mathbbm{1}$ as the identity on all indices. Here we have separated the $U(1)_f \times U(1)_s$ global symmetry into its diagonal part, which couples to $\tilde{A}_1$, and a scalar part which couples to $\tilde{A}_2$. Note that diagonal $U(1)$ global symmetry is ``baryonic.'' Our choice of normalization for the conjugate external field $\tilde{A}_1$ is such that the baryon charges of gauge-invariant operators are integers.

Now consider the five massive phases~\eqref{E:SUphases} obtained by turning on the fermion and scalar masses. The flavor Chern-Simons levels receive one-loop shifts after integrating out the fermions, and now including the flavor groups, the massive phases are described by
\begin{align}
\begin{split}
	\label{E:SUflavor}
	(\text{I}) \, : &\quad SU(N)_{-k+N_f} \times SU(N_f)_{N} \times SU(N_s)_{0} \times J_{\text{I}}\,,
	\\
	(\text{II}) \, : & \quad SU(N)_{-k} \times SU(N_f)_{0}\times SU(N_s)_{0} \times J_{\text{II}}\,,
	\\
	(\text{III}) \, : & \quad SU(N-N_s)_{-k} \times SU(N_f)_{0} \times SU(N_s)_{-k} \times J_{\text{III}}\,,
	\\
	(\text{IVa}) \, : &\quad SU(N-N_s)_{-k+N_f} \times SU(N_f)_{N-N_s} \times SU(N_s)_{-k} \times J_{\text{IVa}}\,,
	\\
	(\text{IVb}) \, : &\quad SU(N-N_s)_{-k+N_f}\times SU(N_f)_{N} \times SU(N_s)_{-k+N_f} \times J_{\text{IVb}}\,,
\end{split}
\end{align}
where again only the first factor is dynamical and the $J$'s refer to $2\times 2$ matrices of abelian Chern-Simons levels. Before computing them, observe that the non-abelian Chern-Simons levels in the phases of the $SU(N)$ theories exactly match the non-abelian levels in the phases of the $U(k)$ theories~\eqref{E:Uflavor}, upon identifying the non-abelian flavor background of the $SU$ theory with that of the $U$ theory,
\beq
	A_{\mu}=A'_{\mu}\,, \qquad B_{\mu} = B_{\mu}'\,.
\eeq

We have to work a bit harder to compute the abelian levels. We find it useful to realize the $SU(N)$ theories as $U(N)\times U(1)$ theories.\footnote{Relatedly, it seems that the simplest way to derive the level/rank duality $SU(N)_{-k} \leftrightarrow U(k)_N$ is to realize the $SU(N)_{-k}$ theory as a suitable $U(N)\times U(1)$ Chern-Simons theory~\cite{Hsin:2016blu}.} First, begin with $U(N)_{-k+\frac{N_f}{2}}$ Chern-Simons theory coupled to $N_f$ fermions and $N_s$ scalars. This theory has a manifest $SU(N_f)\times SU(N_s)\times U(1)_m\times U(1)_F$ global symmetry, where the $U(1)_m$ is a monopole number. Its Lagrangian is
\begin{align}
\begin{split}
	\mathcal{L}_U &= -i \left[ \frac{-k+N_f}{4\pi}\text{tr}\left( ada-\frac{2i}{3}a^3\right) +\frac{1}{2\pi }\text{tr}(a) dA_m+ \frac{N}{4\pi}\text{tr}\left( AdA-\frac{2i}{3}A^3\right) \right]
	\\
	& \qquad + \bar{\psi} i \slashed{D}\psi + |D\phi|^2 + \mathcal{L}_{\rm int}\,,
\end{split}
\end{align}
along with
\begin{align}
\begin{split}
	D_{\mu}\psi & = \left( \partial_{\mu} - i (a_{\mu} \mathbbm{1}_f + A_{\mu}\mathbbm{1}_c - \tilde{A}_{2\mu} \mathbbm{1})\right)\psi\,,
	\\
	D_{\mu}\phi & = \left( \partial_{\mu} - i (a_{\mu} \mathbbm{1}_f + B_{\mu} \mathbbm{1}_c)\right)\phi\,.
\end{split}
\end{align}
Here $a_{\mu}$ is a $U(N)$ gauge field, the other background fields are as before, and $A_m$ is a background field which couples to monopole number. We have also neglected a matrix of abelian background Chern-Simons terms. Now we gauge the monopole number by promoting $A_{m}$ to a dynamical field,
\beq
	A_m \to \tilde{a}\,,
\eeq
and integrating over it in the functional integral. Now there is a new $U(1)$ global symmetry, the monopole number associated with $\tilde{a}$, and we couple said monopole number to a background $U(1)$ field which we take to be $-(\tilde{A}_1 + N \tilde{A}_2)$. The end result is that, with some foresight for the abelian Chern-Simons terms, we redefine the Lagrangian as
\begin{align}
	\nonumber
	\mathcal{L}_U &\to  -i \left[ \frac{-k+N_f}{4\pi}\text{tr}\left( ada-\frac{2i}{3}a^3\right) +\frac{1}{2\pi }\tilde{a}d\left( \text{tr}(a)-(\tilde{A}_1+N\tilde{A}_2)\right) \right] + \bar{\psi} i \slashed{D}\psi + |D\phi|^2	 + \mathcal{L}_{\rm int}
	\\
	\label{E:SUlagrangian2}
	& \qquad \qquad  -i \left[ \frac{N}{4\pi}\text{tr}\left( AdA-\frac{2i}{3}A^3\right) + (k-N_f)\left(\frac{1}{2\pi}\tilde{A}_1d\tilde{A}_2+\frac{N}{4\pi}\tilde{A}_2d\tilde{A}_2\right)\right]\,.
\end{align}
Now $\tilde{a}$ only appears in the Lagrangian through a $BF$ couplings to $\text{tr}(a)$ and so we may integrate it out. This sets the constraint
\beq
	\label{E:constraint}
	\text{tr}(a) = \tilde{A}_1+N \tilde{A}_2\,,
\eeq
which, when inserted back into the Lagrangian, leads to the original $SU(N)$ theory~\eqref{E:SUlagrangian}. In particular, the covariant derivatives of the matter fields now read
\begin{align}
\begin{split}
	D_{\mu}\psi & = \left( \partial_{\mu} - i \left( a_{\mu}\mathbbm{1}_f + A_{\mu} \mathbbm{1}_c + \frac{1}{N}\tilde{A}_{1\mu}\mathbbm{1}\right)\right)\psi\,,
	\\
	D_{\mu}\phi & = \left( \partial_{\mu} - i \left(a_{\mu}\mathbbm{1}_f + B_{\mu}\mathbbm{1}_c + \left(\frac{1}{N}\tilde{A}_{1\mu}+\tilde{A}_{2\mu}\right)\mathbbm{1}\right)\right)\phi\,,
\end{split}
\end{align}
with $a_{\mu}$ a $SU(N)$ gauge field.

Now let us compute the abelian levels. We illustrate the idea in Phase I. In this phase there is no one-loop shift to the levels from integrating out the fermions, and so plugging the constraint~\eqref{E:constraint} into the $U(1)\subset U(N)$ Chern-Simons term in~\eqref{E:SUlagrangian2} leads to the combined abelian Chern-Simons terms
\beq
	-i \left[ \frac{-k+N_f}{4\pi N}(\tilde{A}_1+N \tilde{A}_2)d(\tilde{A}_1 +N \tilde{A}_2) + \frac{k-N_f}{2\pi}\tilde{A}_1d \tilde{A}_2 + \frac{(k-N_f)N}{4\pi} \tilde{A}_2 d\tilde{A}_2 \right] = - i \frac{-k+N_f}{4\pi N}\tilde{A}_1d\tilde{A}_1\,,
\eeq 
i.e. to a matrix of abelian levels
\begin{subequations}
\beq
	\label{E:SUabelian1}
	J_{\text{I}}^{ab} = \begin{pmatrix} \frac{-k+N_f}{N} & 0 \\ 0 & 0 \end{pmatrix} \,,
\eeq
which happily matches the matrix we obtained in Phase I' of the $U(k)$ theory~\eqref{E:Uabelian1}. As for the other phases, a straightforward computation yields
\begin{align}
\begin{split}
	\label{E:SUabelian2}
	J_{\text{II}}^{ab} & = \begin{pmatrix} -\frac{k}{N} & 0 \\ 0 & 0 \end{pmatrix}\,,
	\\
	J_{\text{III}}^{ab} & = -\frac{k}{N-N_s}\begin{pmatrix} 1 & N_s \\ N_s & N N_s \end{pmatrix}\,,
	\\
	J_{\text{IVa}}^{ab} & =  -\frac{k-N_f}{N-N_s}\begin{pmatrix} 1&  N_s \\ N_s & NN_s \end{pmatrix}-N_f N_s \begin{pmatrix}0 & 0 \\ 0 & 1 \end{pmatrix}\,,
	\\
	J_{\text{IVb}}^{ab} & =-\frac{k-N_f}{N-N_s}\begin{pmatrix} 1&  N_s \\ N_s & NN_s \end{pmatrix}\,.
\end{split}
\end{align}
\end{subequations}
These matrices precisely match those computed in the corresponding Phases~\eqref{E:Uabelian2} of the $U(k)$ theory, provided that we identify the external $U(1)$ fields as
\beq
\tilde{A}_1 = \tilde{A}_1' \,, \qquad \tilde{A}_2 = \tilde{A}_2'\,.
\eeq
Observe that the baryonic symmetry of the $SU(N)$ theory, which coupled to $\tilde{A}_1$, is mapped to the monopole symmetry of the $U(k)$ theory, which coupled to $\tilde{A}_1'$. Taken together, we see that all flavor Chern-Simons terms can be matched across the duality.

\section{Global symmetries}
\label{S:global}

The point of this Section is two-fold. We have already seen that the $SU/U$ duality exchanges the baryon number of the $SU$ theory with the monopole number of the $U$ theory. In the next Subsection we see how this works in more detail, matching the quantum numbers of baryons with those of the monopoles. In Subsection~\ref{S:exactSymmetries} we deduce the faithful subgroup of the manifest $SU(N_f)\times U(N_s)\times U(1)$ global symmetry that acts on both sides of the duality, finding that this faithful global symmetry matches. Our discussion closely imitates that of~\cite{Aharony:2015mjs,Benini:2017dus}. 

\subsection{Baryons and monopoles}

In the last Section we parameterized the $U(1)_f \times U(1)_s$ global symmetry of the $SU(N)$ theories with some foresight. We rewrote it in terms of a $U(1)_b \times U(1)_F$ global symmetry, where the first factor is ``baryonic'' and the second is an ordinary global symmetry. Under them the fundamental fermions $\psi$ and scalars $\phi$ have charges
\beq
\begin{tabular}{|c|c|c|}
	\hline
	 & $U(1)_b$ & $U(1)_F$ \\
	 \hline
	 $\psi$ & $\frac{1}{N}$ & $0$ \\
	 \hline
	 $\phi$ & $\frac{1}{N}$ & $1$ \\
	 \hline
 \end{tabular}
\eeq
For simplicity, we take the large $N$ and large $k$ limit with $N/k$ fixed. We further take $N_f = N_s=1$, although it is straightforward to allow for a more flavors.

 The various gauge-invariant operators of the $SU(N)$ theory fall into two types. The first are the mesons 
\beq
\bar{\psi} D^n \psi\,, \qquad D^m\bar{\psi} \cdot D^n \phi\,, \qquad \phi^{\dagger}D^n\phi\,.
\eeq
These operators remain ``light'' at large $N$ with a dimension that scales as $O(N^0)$. They all have zero baryon charge and only the second type is charged under $U(1)_F$. There are also ``multi-trace'' operators built out of products of the mesons and derivatives. The second class of operators are baryons. The simplest baryons are a product of $N$ fundamental fermions and bosons with the color indices antisymmetrized. Our convention is that they carry charge $+1$ under $U(1)_b$. There must be derivatives acting on the scalars in order to antisymmetrize them, so, schematically the baryons are
\beq
	\label{E:baryon}
	\varepsilon \underbrace{\psi \hdots \psi}_{N-M} \underbrace{ \phi D \phi \hdots D^n\phi}_{M}\,,
\eeq
Besides carrying charge $+1$ under baryon number, they also carry charge $+M$ under $U(1)_F$. There are many such baryons, depending on how we take derivatives. A simple counting exercise at large $M$ with $M/N$ fixed~\cite{Shenker:2011zf} reveals that the dimension of the lowest-dimension baryons at large $N$ is approximately given by $N-M + \frac{2}{3}M^{\frac{3}{2}}$. Observe that this is the dimension of a baryon in $SU(M)$ theory with a fermion plus that of a baryon in $SU(N-M)$ theory with a scalar. There are also multi-trace operators built out of products of the simplest baryons, mesons, and derivatives.

Now for the $U(k)$ theories. In these theories there is a $U(1)\times U(1)$ global symmetry under which the various fields are charged as
\beq
\begin{tabular}{|c|c|c|}
	\hline
	 & $U(1)_m$ & $U(1)_F$ \\
	 \hline
 	$\Phi$ & 0 & 0 \\
	\hline
	$\Psi$ & 0 & -1 \\
	\hline
\end{tabular}
\eeq
All fundamental fields are neutral under the monopole number $U(1)_m$, and instead the monopole current is given by the $U(1)\subset U(k)$ gauge field, $j_m^{\mu} = \frac{1}{2\pi}\varepsilon^{\mu\nu\rho}\partial_{\nu}\text{tr}(a'_{\rho})$. As in the $SU(N)$ theories there are mesons
\beq
\Phi^{\dagger} D^n \Phi \,, \qquad D^m \Phi \cdot D^n \bar{\Psi}\,, \qquad \bar{\Psi}D^n\Psi\,,
\eeq
which remain light at large $N$ with a dimension that scales as $O(N^0)$. All of these operators carry zero monopole number and only the second is charged under $U(1)_F$. The second set of operators are monopoles. For a $U(k)$ gauge theory these are characterized by a set of $k$ integers $q_i$ (up to permutations by the Weyl group) which give the $U(1)^k\subset U(k)$ fluxes. These integers are called GNO charges, and by our convention the total monopole charge is $\sum_i q_i$. Monopole operators are not gauge-invariant in $U(k)_N$ Chern-Simons theories: in the presence of a monopole with $n$ GNO charges the gauge symmetry is effectively broken to $U(1)^n\times U(k-n)$, and due to the Chern-Simons term the monopole carries charge $N q_i$ under the $i^{th}$ $U(1)$. These must be canceled by inserting additional fields in the (anti-)fundamental representation of $U(k)$ so as to obtain a gauge-invariant operator. For example, the simplest monopoles carry GNO charges $\{q_i\} = \{1,0,\hdots,0\}$, and these are expected to have the lowest dimension of any monopoles. To make the monopole gauge-invariant we must multiply it by $N$ anti-fundamental fields with the same gauge index. We can make up such an operator out of $M$ scalars and $N-M$ fermions, but to do so we must symmetrize the fermions by including appropriate derivatives. So, schematically, these monopoles take the form
\beq
	\label{E:monopole}
	(\text{GNO flux}) \times \underbrace{\Phi^{\dagger} \hdots \Phi^{\dagger}}_{N-M} \underbrace{ \bar{\Psi} D \bar{\Psi} \hdots D^n \bar{\Psi}}_{M}\,.
\eeq
Clearly these operators carry monopole number $+1$ as well as charge $+M$ under $U(1)_F$, which coincides with the $U(1)_b\times U(1)_F$ charges of the baryons in~\eqref{E:baryon}. At large $M$ with $M/N$ fixed, the dimension of the lowest-dimension monopoles are given by the same counting argument as for the baryons of the $SU(N)$ theory, with $\Delta = N-M + \frac{2}{3}M^{\frac{3}{2}}$. There are many such operators with various spins. In the monopole background the scalars carry spin$-1/2$~\cite{Wu:1976ge} while the fermions are spin-0, so that the possible spin quantum numbers of the monopoles precisely equals the set of possible spin quantum numbers for the baryons of the $SU(N)$ theory. In sum, at large $N$, the quantum numbers of the simplest baryons~\eqref{E:baryon} in the $SU$ theory match those of the simplest monopoles~\eqref{E:monopole} in the $U(k)$ theory.

\subsection{Exact flavor symmetries}
\label{S:exactSymmetries}

Let us work out the faithfully acting global symmetries that act on the local operators on both sides of our proposed duality~\eqref{E:proposal} for generic values of the parameters.

The $SU(N)$ theories have a naive $U(N_f)\times U(N_s)$ global symmetry, where the $U(N_f)$ acts on the $N_f$ fundamental fermions and the $U(N_s)$ on the fundamental scalars. However only a $\left( U(N_f)\times U(N_s)\right)/\mathbb{Z}_N$ subgroup of this symmetry acts faithfully on the operator spectrum, where the generator of $\mathbb{Z}_N$ acts as multiplication by $e^{2\pi i/N}$. In physical terms, the gauge-invariant operators charged under the diagonal $U(1)$ symmetry are baryons built from $N$ fundamental fermions and bosons. There is also a charge conjugation symmetry $\mathbb{Z}_2^C$ which exchanges the fundamental representation with the anti-fundamental representation, so that the total global symmetry is
\beq
	\label{E:SUsymmetry}
	\faktor{\big( U(N_f)\times U(N_s)\big)}{\mathbb{Z}_N} \rtimes \mathbb{Z}_2^C\,.
\eeq

What is the faithful global symmetry that acts on the $U(k)$ theories? Here there is a manifest $SU(N_f)\times U(N_s) \times U(1)_m$ global symmetry where $U(1)_m$ is monopole number. As we discussed above, monopoles are characterized by a set of GNO fluxes $\{q_i\}$ and the total monopole charge is their sum $q_m = \sum_i q_i$. Monopoles carry electric charge by virtue of the bare $U(k)_N$ Chern-Simons term, so to render a monopole gauge-invariant it must be dressed with a number of fundamental and anti-fundamental fields. The total number of anti-fundamental fields minus the number of fundamentals must equal $Nq_m$. Given monopoles with $U(1)_F$ charge $M$, the monopoles fill out representations of $SU(N_f)\times SU(N_s) $ with $N_f$-ality $(Nq_m -M)\text{ mod } N_f$ and $N_s$-ality $M \text{ mod }N_s$. We may then understand $U(1)_m$ to act as a diagonal $U(1)$, reducing $SU(N_f)\times U(N_s)\times U(1)_m$ to $U(N_f)\times U(N_s)$, subject to an additional $\mathbb{Z}_N$ quotient. As in the $SU(N)$ theories there is also a $\mathbb{Z}_2^C$ charge conjugation symmetry, so that the total faithfully acting global symmetry is
\beq
	\faktor{\big( U(N_f)\times U(N_s)\big) }{\mathbb{Z}_N} \rtimes \mathbb{Z}_2^C\,,
\eeq	
which precisely matches~\eqref{E:SUsymmetry}. 

In this work we do not undertake an analysis of the quantization conditions for flavor Chern-Simons terms consistent with the faithfully acting global symmetry~\eqref{E:SUsymmetry}. For typical values of the parameters it will be the case that those flavor Chern-Simons terms will necessarily have levels with a fractional part. When it exists this fractional part is an 't Hooft anomaly, and it implies that these theories do not have an intrinsically 3d definition in a general flavor background, and must instead be defined as living on the boundary of a 4d SPT phase. While we do not deduce the anomalies of the theories in our proposed duality~\eqref{E:proposal}, we do observe that since the global symmetries match, as do the flavor Chern-Simons terms in the various phases, then whatever the 't Hooft anomalies are in the $SU/U$ theories, they should match.

\section{$SU/U$ duality implies a $U/U$ duality}
\label{S:Uduality}

The basic sequences of 3d bosonization dualities equate
\begin{align*}
	SU(N)_{-k+\frac{N_f}{2}} \quad \text{ with }N_f \text{ fermions }\quad 
	&\leftrightarrow 
	\quad U(k)_N \quad \text{ with } N_f \, \text{ scalars}\,,
	\\
	U(N)_{-k+\frac{N_f}{2}} \quad \text{with } N_f \text{ fermions } \quad
	& \leftrightarrow
	\quad SU(k)_{N} \quad \text{ with }N_f \text{ scalars}\,,
\end{align*}
for $N_f \leq k$. There are also time-reversed versions of these conjectures. These dualities are in fact equivalent to each other and yet another duality,
\beq
	\label{E:sequence3}
	U(N)_{-k+\frac{N_f}{2}, -k+\frac{N_f}{2}\pm N} \, \text{ with }N_f \text{ fermions } \,\,
	\leftrightarrow
	\,\, U(k)_{N,N\mp k} \,\, \text{ with } N_f \text{ scalars}\,.
\eeq
The two subscripts denote independent levels for the non-abelian and Abelian parts of $U(N)$,
\beq
U(N)_{k,k+mN} = (SU(N)_k \times U(1)_{kN +mN^2})/\mathbb{Z}_N\,.
\eeq
To see that these dualities are equivalent to each other, start with the first duality, defining both sides with a suitable and matched choice of background Chern-Simons term for the $U(1)$ global symmetry. Then gauge the $U(1)$ global symmetry on both sides. The same procedure with a slightly different choice of background $U(1)$ Chern-Simons level on both sides gives the $U/U$ duality. A completely general choice of $U(1)$ Chern-Simons level leads to yet more dualities between $U(N)$ theories and $U(k)\times U(1)$ theories where the $U(1)$ factor is topological~\cite{Radicevic:2016wqn}. 

As a side comment, the massive phases of the $U/U$ duality~\eqref{E:sequence3} match by virtue of a level/rank duality for $U$ gauge groups (see~\cite{Hsin:2016blu}),
\beq
	\label{E:levelrank2}
	U(N)_{-k,-k\pm N} \quad \leftrightarrow \quad U(k)_{N,N\mp k}\,,
\eeq
which also follows from simply setting $N_f = 0$ in~\eqref{E:sequence3}.

Using the same sort of logic, our proposed duality~\eqref{E:proposal} implies a $U/U$ duality between
\beq
	\label{E:proposal2}
	U(N)_{-k+\frac{N_f}{2},-k+\frac{N_f}{2}\pm N} \, \text{ with } N_f\, \psi\,, \, N_s\,  \phi
	\,\,\,\,\,\leftrightarrow\,\,\,\,\,
	U(k)_{N-\frac{N_s}{2},N-\frac{N_s}{2}\mp k} \, \text{ with }N_s\, \Psi\,, \, N_f \, \Phi\,.
\eeq
Both sides have a phase diagram that looks identical to those of the $SU(N)$ and $U(k)$ theories discussed in Section~\ref{S:phase}, and it is easy to see that the massive phases and critical lines all match on account of the level/rank duality~\eqref{E:levelrank2} and $U/U$ bosonization duality~\eqref{E:sequence3}. We also require that the quartic operator $(\bar{\psi}\cdot \phi)(\phi^{\dagger}\cdot \psi)$ be present on both sides of the duality with the same sign as the non-abelian Chern-Simons level.

We start with the Lagrangians for both sides of our proposed $SU(N)/U(k)$ duality, matched so that all flavor Chern-Simons levels agree in massive phases. For simplicity, we only turn on a background $\tilde{A}_1$ for the baryon/monopole $U(1)$ global symmetry. With a useful convention for the bare $U(1)$ Chern-Simons level, we have
\begin{align}
\begin{split}
	\label{E:beforeUU}
	\mathcal{L}_{SU(N)} & = -i  \left[ \frac{-k+N_f}{4\pi}\text{tr}\left( a da-\frac{2i}{3}a^3\right)  + \frac{-k+N_f}{4\pi N}\tilde{A}_1d\tilde{A}_1\right]+ \bar{\psi} i \slashed{D} \psi + |D\phi|^2 + \mathcal{L}_{\rm int} \,,
	\\
	\mathcal{L}_{U(k)} & = -i  \left[ \frac{N}{4\pi}\text{tr}\left( a' da' -\frac{2i}{3}a'^3\right)+\frac{1}{2\pi}\text{tr}(a') d\tilde{A}_1\right] +|D\Phi|^2 + \bar{\Psi}i \slashed{D} \Psi + \mathcal{L}_{\rm int}'\,.
\end{split}
\end{align}
Here
\begin{align*}
	D_{\mu}\psi & = \left( \partial_{\mu} - i \left(a_{\mu}\mathbbm{1}_f + \frac{1}{N}\tilde{A}_{1\mu}\mathbbm{1}\right)\right)\psi\,,
	\\
	D_{\mu}\phi & = \left( \partial_{\mu} - i \left( a_{\mu}\mathbbm{1}_f + \frac{1}{N}\tilde{A}_{1\mu}\right)\right)\phi\,,
	\\
	D_{\mu}\Phi & = \left( \partial_{\mu} - i  a'_{\mu}\mathbbm{1}_f\right)\Phi\,,
	\\
	D_{\mu}\Psi & = \left( \partial_{\mu} - i a'_{\mu} \mathbbm{1}_f \right) \Psi\,.
\end{align*}
Before going on, observe that if we promote $\tilde{A}_1$ to be a dynamical gauge field,
\beq
\tilde{A}_1 \to \tilde{a}\,,
\eeq
then in the $SU(N)$ theory it combines with $a$ into a $U(N)$ gauge field $\bar{a} = a +\frac{\tilde{a}}{N}\mathbbm{1}$ so that the $SU(N)$ theory becomes $U(N)_{-k+\frac{N_f}{2}}$ Chern-Simons theory coupled to $N_f$ scalars and $N_s$ scalars,
\beq
	\mathcal{L}_{SU(N)}\to \mathcal{L}_{U(N)}  = -i \frac{-k+N_f}{4\pi}\text{tr}\left( \bar{a}d\bar{a} - \frac{2i}{3}\bar{a}^3\right) + \bar{\psi} i \slashed{D}\psi + |D\phi|^2 + \mathcal{L}_{\rm int} \,.
\eeq
Meanwhile in the $U(k)$ theory, integrating over $\tilde{a}$ enforces $\text{tr}(a') = 0$ turning it into $SU(k)_{N-\frac{N_s}{2}}$ Chern-Simons theory coupled to $N_f$ scalars and $N_s$ fermions,
\beq
	\mathcal{L}_{U(k)} \to \mathcal{L}_{SU(k)} = - i \frac{N}{4\pi}\text{tr}\left( \bar{a}'d\bar{a}' - \frac{2i}{3}\bar{a}'^3\right) + |D\Phi|^2 + \bar{\Psi} i \slashed{D} \Psi + \mathcal{L}_{\rm int}'\,.
\eeq 
So we see that the original $SU(N)/U(k)$ duality implies a $U(N)/SU(k)$ duality
\beq
	U(N)_{-k+\frac{N_f}{2}} \, \text{ with } N_f\, \psi\,, \, N_s \, \phi \quad
	\leftrightarrow
	\quad SU(k)_{N-\frac{N_s}{2}} \, \text{ with } N_s \, \Psi\,, \, N_f \, \Phi \,.
\eeq
This is merely the time-reversed version of the original $SU(N)/U(k)$ duality combined with exchanging $N\leftrightarrow k$, $N_f \leftrightarrow N_s$. This is yet another consistency check on our proposal~\eqref{E:proposal}.

We can now obtain the $U/U$ dualities~\eqref{E:proposal2}. To the $SU(N)$ and $U(k)$ Lagrangians in~\eqref{E:beforeUU} we now add a background Chern-Simons term with level $\pm 1$ for the $U(1)$ baryon/monopole global symmetry,
\begin{align}
\begin{split}
	\mathcal{L}_{SU(N)} &\to \mathcal{L}_{SU(N)} \pm -i\frac{1}{4\pi} \tilde{A}_1d\tilde{A}_1\,,
	\\
	\mathcal{L}_{U(k)} & \to \mathcal{L}_{U(k)} \pm -i \frac{1}{4\pi}\tilde{A}_1d\tilde{A}_1\,.
\end{split}
\end{align}
Now, we promote $\tilde{A}_1$ to a dynamical gauge field $\tilde{A}_1 \to \tilde{a}$. On the $SU(N)$ side it combines with $a$ into a $U(N)$ gauge field $\bar{a} = a + \frac{\tilde{a}}{N}\mathbbm{1}$ and the extra $U(1)$ Chern-Simons term shifts the $U(1)$ level by $\pm N$, giving
\beq
	SU(N)_{-k+\frac{N_f}{2}} \, \text{ with } N_f \, \psi\,, \, N_s \, \phi \quad \to \quad U(N)_{-k+\frac{N_f}{2},-k+\frac{N_f}{2}\pm N} \, \text{ with }N_f \, \psi\,, \, N_s \, \phi\,,
\eeq
which is the left side of the $U/U$ duality~\eqref{E:proposal2}. In the $U(k)$ theory, the new field $\tilde{a}$ appears in two terms:
\beq
	\mathcal{L}_{\tilde{a}} = -i \left[ \frac{1}{2\pi}\text{tr}(a)d\tilde{a} \pm \frac{1}{4\pi}\tilde{a}d\tilde{a}\right]\,.
\eeq
It can be integrated out, giving
\beq
\tilde{a} = \mp \text{tr}(a)\,,
\eeq
so that
\beq
	\mathcal{L}_{\tilde{a}} \to -i \frac{\mp 1}{4\pi}\text{tr}(a) d\text{tr}(a)\,,
\eeq
which effectively shifts the $U(1)$ level of the $U(k)$ Chern-Simons term by $\mp k$. In this way the $U(k)_{N-\frac{N_s}{2}}$ theory becomes
\beq
	U(k)_{N-\frac{N_s}{2}} \, \text{ with }N_s \, \Psi\,, \, N_f \, \Phi \quad \to \quad U(k)_{N-\frac{N_s}{2},N-\frac{N_s}{2}\mp k} \, \text{ with } N_s \, \Psi\,, \, N_f \, \Phi\,,
\eeq
which is the right side of the $U/U$ duality~\eqref{E:proposal2}. So the $SU/U$ duality~\eqref{E:proposal} implies the $U/U$ duality~\eqref{E:proposal2} as promised.

\section{$SO$ and $USp$ dualities}
\label{S:real}

Another infinite sequence of level/rank dualities equates~\cite{Aharony:2016jvv}
\begin{align}
\begin{split}
	\label{E:levelrank3}
	SO(N)_{-k} \quad & \leftrightarrow \quad SO(k)_N\,,
	\\
	USp(2N)_{-k} \quad & \leftrightarrow \quad USp(2k)_N \,.
\end{split}
\end{align}
One might expect that there are ``flavored'' versions of these dualities, and indeed there is a natural conjecture for them~\cite{Aharony:2016jvv,Metlitski:2016dht}:
\begin{align}
\begin{split}
	\label{E:sequence4}
	SO(N)_{-k+\frac{N_f}{2}} \, \text{ with } N_f \, \text{ real fermions} \quad
	& \leftrightarrow
	\quad SO(k)_N \, \text{ with }N_f \, \text{ real scalars} \,,
	\\
	USp(2N)_{-k+\frac{N_f}{2}} \, \text{ with } N_f \, \text{ fermions} \quad 
	& \leftrightarrow
	\quad USp(2k)_N \, \text{ with } N_f\, \text{ scalars}\,.
\end{split}
\end{align}
As before, on the scalar side there is a non-trivial scalar potential tuned to criticality so that the scalars are (real) WF scalars, while the fermions are Majorana. There are also flavor bounds $N_f\leq k$ $USp$ for the dualities, while for the $SO$ dualities one requires $N_f \leq k-2$ for $N=1$, $N_f \leq k-1$ for $N=2$, and $N_f\leq k$ for $N>2$. Equivalently, the $SO$ dualities simultaneously require $N_f \leq k$ and $3+N_f \leq k+N$.

The evidence for these dualities is at the same level as for the basic sequence of $SU/U$ bosonization dualities~\eqref{E:sequence1}. To leading order in large $N$ the orthogonal and symplectic theories are just orbifolds of the $SU/U$ theories~\cite{Aharony:2016jvv}. The massive phases of both sides of the dualities match, as do the exact global symmetries and 't Hooft anomalies~\cite{Benini:2017dus}.

It is then natural to conjecture a sequence of $SO/SO$ and $USp/USp$ bosonization dualities with both fundamental fermions and bosons. We propose
\begin{align}
	\label{E:proposalSO}
	SO(N)_{-k+\frac{N_f}{2}} \, \text{ with } N_f \, \psi\,, \, N_s \, \phi \quad
	& \leftrightarrow
	\quad SO(k)_{N-\frac{N_s}{2}} \, \text{ with }N_s\, \Psi\,, \, N_f\, \Phi\,,
	\\
	\label{E:proposalSp}
	USp(2N)_{-k+\frac{N_f}{2}} \, \text{ with } N_f\, \psi\,, \, N_s \, \phi \quad
	& \leftrightarrow
	\quad USp(2k)_{N-\frac{N_s}{2}} \, \text{ with }N_s \, \Psi\,, \, N_f\, \Phi\,.
\end{align}
For the $USp$ dualities we require $N_f\leq k$ and $N_s\leq N$, and for the $SO$ dualities we further require $3+N_s+N_f \leq k+N$. At large $N$ these dualities follow from our original $SU/U$ conjecture~\eqref{E:proposal} by suitable orbifold projections. 

Under the assumption that the only relevant operators in these theories are scalar and fermion mass operators, we may proceed just as in Section~\ref{S:phase} and map out the phase diagrams of the dual pairs. For general parameters, the phase diagrams look identical to those of the $SU/U$ theories in Figure~\ref{F:phases}, the massive phases match on account of the level/rank dualities~\eqref{E:levelrank3}, and the critical lines match on account of~\eqref{E:sequence4}. For the orthogonal sequence, the matching of the critical lines requires the flavor bound $3+N_s+N_f \leq k+N$.

Let us obtain this last flavor bound, starting with the $SO(N)$ theories. As in our discussion of the $SU/U$ dualities, we require the operator $(\bar{\psi}^i \cdot \phi_m)(\phi^m \psi_i)$ to be present with a coefficient with the same sign as the Chern-Simons level. For general parameters, there are then five distinct phases and four different TFTs, given by
\begin{align}
\begin{split}
	\label{E:SONphases}
	(\text{I}) \quad m_{\psi}>0\,, \, m_{\phi}^2>0 \,: &\quad SO(N)_{-k+N_f} \,,
	\\
	(\text{II}) \quad m_{\psi}<0\,, \, m_{\phi}^2>0 \,: &\quad SO(N)_{-k}\,,
	\\
	(\text{III}) \quad m_{\psi}<0\,, \, m_{\phi}^2<0 \,: &\quad SO(N-N_s)_{-k}\,,
	\\
	(\text{IV}) \quad m_{\psi}>0\,, \, m_{\phi}^2<0\, : & \quad SO(N-N_s)_{-k+N_f}\,.
\end{split}
\end{align}
Phase IV splits into two, 
\begin{align}
\begin{split}
	(\text{IVa}) \quad m_{\psi}>0\,, \, m_s <0 \,, \, m_{\phi}^2<0 \,: & \quad SO(N-N_s)_{-k+N_f}\,,
	\\
	(\text{IVb}) \quad m_{\psi}>0\,, \, m_s>0 \,, \, m_{\phi}^2 <0 \, : & \quad SO(N-N_s)_{-k+N_f}\,.
\end{split}
\end{align}
The critical lines are described by
\begin{align}
\begin{split}
	\label{E:SONlines}
	(\text{I-II}) \quad &SO(N)_{-k+\frac{N_f}{2}} \text{ with }N_f \, \psi\,,
	\\
	(\text{II-III}) \quad &SO(N)_{-k} \text{ with }N_s\, \phi\,,
	\\
	(\text{III-IVa}) \quad &SO(N-N_s)_{-k+\frac{N_f}{2}} \text{ with }N_f\, \psi\,,
	\\
	(\text{IVa-IVb}) \quad & N_s N_f \text{ singlet } \psi\text{'s} \, + \, SO(N-N_s)_{-k+N_f} \text{ TFT}\,,
	\\
	(\text{IVb-I}) \quad & SO(N)_{-k+N_f} \text{ with }N_s \,\phi\,.
\end{split}
\end{align}
The corresponding phase diagram is identical to that on the left side of Figure~\ref{F:phases}. 

Meanwhile, there are generally five phases of the $SO(k)$ theories described by four different TFTs,
\begin{align}
\begin{split}
	\label{E:SOkphases}
	(\text{I'}) \quad m_{\Psi}>0\,, \, m_{\Phi}^2<0 \,: &\quad SO(k-N_s)_{N} \,,
	\\
	(\text{II'}) \quad m_{\Psi}>0\,, \, m_{\Phi}^2>0 \,: &\quad SO(k)_{N}\,,
	\\
	(\text{III'}) \quad m_{\Psi}<0\,, \, m_{\Phi}^2>0 \,: &\quad SO(k)_{N-N_s}\,,
	\\
	(\text{IV'}) \quad m_{\Psi}<0\,, \, m_{\Phi}^2<0\, : & \quad SO(k-N_f)_{N-N_s}\,.
\end{split}
\end{align}
We require the coefficient of the $(\bar{\Psi}^m\cdot \Phi_i)(\Phi^i\cdot \Psi_m)$ operator to be nonzero and positive, so that Phase IV' splits into two,
\begin{align}
\begin{split}
	(\text{IVa'}) \quad m_{\Psi}<0\,, \, m_s <0 \,, \, m_{\phi}^2<0 \,: & \quad SO(k-N_f)_{N-N_s}\,,
	\\
	(\text{IVb'}) \quad m_{\psi}<0\,, \, m_s>0 \,, \, m_{\phi}^2 <0 \, : & \quad SO(k-N_f)_{N-N_s}\,.
\end{split}
\end{align}
The critical lines are
\begin{align}
\begin{split}
	\label{E:SOklines}
	(\text{I'-II'}) \quad &SO(k)_{N} \text{ with }N_f \, \Phi\,,
	\\
	(\text{II'-III'}) \quad &SO(k)_{N-\frac{N_s}{2}} \text{ with }N_s\, \Psi\,,
	\\
	(\text{III'-IVa'}) \quad &SO(k)_{N-N_s} \text{ with }N_f\, \Phi\,,
	\\
	(\text{IVa'-IVb'}) \quad & N_s N_f \text{ singlet } \Psi\text{'s} \, + \, SO(k-N_f)_{N-N_s} \text{ TFT}\,,
	\\
	(\text{IVb'-I'}) \quad & SO(k-N_f)_{N-\frac{N_s}{2}} \text{ with }N_s \,\Psi\,,
\end{split}
\end{align}
and the phase diagram coincides with the right side of Figure~\ref{F:phases}.

The massive phases of the $SO(N)$ theory~\eqref{E:SONphases} clearly match those of the $SO(k)$ theory~\eqref{E:SOkphases} upon using the $SO$ level/rank duality~\eqref{E:levelrank3}. Similarly the theories arising on the critical lines match by virtue of the $SO$ bosonization duality~\eqref{E:sequence4}, and for the case of the IVa-IVb and IVa'-IVb' lines one must also use the $SO$ level/rank duality. However, the flavor bounds on the $SO$ dualities~\eqref{E:sequence4} $N_f \leq k$ and $3+N_f \leq k+N$ only hold for lines III-IVa and III'-IVa' if
\beq
3+N_s + N_f \leq k+N\,,
\eeq
is satisfied, which originates that bound.

We leave a careful computation of the exact global symmetries and their 't Hooft anomalies for future work.

\section{Conclusion}
\label{S:conclude}

In this work we have conjectured new infinite sequences of dualities between non-supersymmetric Chern-Simons-matter theories with fundamental bosons and fermions. The three inequivalent dualities are the ``basic'' $SU/U$ sequence~\eqref{E:proposal}, a real $SO/SO$ sequence~\eqref{E:proposalSO}, and a $USp/USp$ sequence~\eqref{E:proposalSp}. We performed some basic consistency checks on our proposal, including the matching of phase diagrams and, for the $SU/U$ dualities, the matching of global symmetries.

We conclude with a short list of open questions.

\begin{enumerate}
	\item Komargodski and Seiberg have recently suggested~\cite{Komargodski:2017keh} that the basic bosonization duality~\eqref{E:sequence1} and its cousins may be extended beyond the flavor bound $N_f \leq k$. Their proposal is that new ``quantum'' phases open up in the range $k< N_f< N_*(N,k)$ where $N_*(N,k)$ is some presently unknown function of $N$ and $k$. Their logic has also been useful in a proposal to map out the phase diagram of Chern-Simons theory with a single adjoint fermion~\cite{Gomis:2017ixy}. It would be interesting to understand to what extent our conjecture~\eqref{E:proposal} can also be extended beyond the flavor bounds $N_f \leq k$ and $N_s \leq N$. 
	\item There are a host of supersymmetric dualities between 3d field theories with at least $\mathcal{N}=2$ supersymmetry (SUSY), including Seiberg duality (sometimes called Giveon/Kutasov duality~\cite{Giveon:2008zn} in three dimensions) and mirror symmetry~\cite{Intriligator:1996ex}. At large $N$ there is significant evidence that the basic 3d bosonization dualities are inherited from a Seiberg duality between certain $\mathcal{N}=2$ supersymmetric Chern-Simons theories with unitary gauge group coupled to chiral multiplets~\cite{Jain:2013gza,Gur-Ari:2015pca} (although it seems~\cite{upcoming} that the particular flow studied in~\cite{Jain:2013gza} does not work as advertised). After turning on a deformation which completely breaks SUSY, the low-energy theory on both sides of the duality flows to a product of bosonic and fermionic Chern-Simons-matter theories. The underlying Seiberg duality then exchanges the bosonic half of the electric theory with the fermionic part of the magnetic one, and the fermionic half of the electric theory with the bosonic part of the magnetic one. 
	
	However there are more general Giveon/Kutasov dualities with a single gauge group as studied in~\cite{Benini:2011mf}. These equate a $\mathcal{N}=2$ SUSY-Chern-Simons theory with $N_f$ chiral and $N_f'$ anti-chiral multiplets (meaning matter in both the fundamental and anti-fundamental representations) and a $SU(N_f)\times SU(N_f')$ global symmetry with another $\mathcal{N}=2$ theory with chiral and anti-chiral multiplets and, reminiscent of 4d Seiberg duality, gauge-neutral mesons which are bifundamental under the flavor symmetry. It would be nice to understand if our proposed duality~\eqref{E:proposal} is inherited from these more general dualities along the lines of~\cite{Jain:2013gza,Gur-Ari:2015pca}, and, if so, to understand what happens to the mesons. (Indeed one of the original motivations behind the present work was to identify a non-SUSY bosonization duality with mesons, although, as we found, a duality with fermions and bosons does not seem to be allow for such gauge-neutral fields.)
	\item Chern-Simons theory with fundamental matter is analytically tractable in the large $N$ and $k$ limit with $N/k$ fixed, and indeed, the best evidence for the non-supersymmetric bosonization dualities comes from explicit computations of correlation functions, the thermal partition function, and scattering amplitudes in that limit. It would be nice to extend those computations to allow for multiple bosons and fermions. In particular, one ought to be able to address the perplexing questions related to the to the quartic operator $(\bar{\psi}\cdot \phi)(\phi^{\dagger}\cdot \psi)$ which played an important role in our proposed duality. In Section~\ref{S:phase} we saw that if this operator was tuned away, or if its coefficient had the wrong sign, then the duality~\eqref{E:proposal} was inconsistent. But this sign (or vanishing) is likely determined by the underlying dynamics, which are yet unsolved. Relatedly, it is not yet known if this operator is irrelevant (in which case it is in fact dangerously irrelevant) or relevant at large but finite $N$. We intend to return to these questions soon.
	\item Finally, recall the conjectured duality between the Chern-Simons theories with one boson or one fermion at large $N, k$ with $N/k$ finite. A natural question is then if there is a Vasiliev-like theory dual to Chern-Simons theory with both $N_f$ fermions and $N_s$ scalars, and if so, if our proposed duality is consistent with it.
\end{enumerate}

\section*{Acknowledgements}

We are especially grateful to A.~Karch for many enlightening conversations as well as for initial collaboration on this project. We would also like to thank O.~Aharony, K.~Aitken, Z.~Komargodski, and R.~Yacoby for useful comments, as well as the Simons Center for Geometry and Physics for hospitality while a portion of this project was completed. This work was supported in part by the US Department of Energy under grant number DE-SC0013682.

\bibliographystyle{JHEP}
\bibliography{master}

\providecommand{\href}[2]{#2}\begingroup\raggedright\begin{thebibliography}{10}

\bibitem{Naculich:1990pa}
S.~G. Naculich, H.~A. Riggs, and H.~J. Schnitzer, {\it {Group Level Duality in
  {WZW} Models and {Chern-Simons} Theory}},  {\em Phys. Lett.} {\bf B246}
  (1990) 417--422.

\bibitem{Mlawer:1990uv}
E.~J. Mlawer, S.~G. Naculich, H.~A. Riggs, and H.~J. Schnitzer, {\it {Group
  level duality of WZW fusion coefficients and Chern-Simons link observables}},
   {\em Nucl. Phys.} {\bf B352} (1991) 863--896.

\bibitem{Nakanishi:1990hj}
T.~Nakanishi and A.~Tsuchiya, {\it {Level rank duality of WZW models in
  conformal field theory}},  {\em Commun. Math. Phys.} {\bf 144} (1992)
  351--372.

\bibitem{Aharony:2011jz}
O.~Aharony, G.~Gur-Ari, and R.~Yacoby, {\it {d=3 Bosonic Vector Models Coupled
  to Chern-Simons Gauge Theories}},  {\em JHEP} {\bf 03} (2012) 037,
  \href{http://xxx.lanl.gov/abs/1110.4382}{{\tt 1110.4382}}.

\bibitem{Giombi:2011kc}
S.~Giombi, S.~Minwalla, S.~Prakash, S.~P. Trivedi, S.~R. Wadia, and X.~Yin,
  {\it {Chern-Simons Theory with Vector Fermion Matter}},  {\em Eur. Phys. J.}
  {\bf C72} (2012) 2112, \href{http://xxx.lanl.gov/abs/1110.4386}{{\tt
  1110.4386}}.

\bibitem{Aharony:2012nh}
O.~Aharony, G.~Gur-Ari, and R.~Yacoby, {\it {Correlation Functions of Large N
  Chern-Simons-Matter Theories and Bosonization in Three Dimensions}},  {\em
  JHEP} {\bf 12} (2012) 028, \href{http://xxx.lanl.gov/abs/1207.4593}{{\tt
  1207.4593}}.

\bibitem{Aharony:2015mjs}
O.~Aharony, {\it {Baryons, monopoles and dualities in Chern-Simons-matter
  theories}},  {\em JHEP} {\bf 02} (2016) 093,
  \href{http://xxx.lanl.gov/abs/1512.00161}{{\tt 1512.00161}}.

\bibitem{Hsin:2016blu}
P.-S. Hsin and N.~Seiberg, {\it {Level/rank Duality and Chern-Simons-Matter
  Theories}},  {\em JHEP} {\bf 09} (2016) 095,
  \href{http://xxx.lanl.gov/abs/1607.07457}{{\tt 1607.07457}}.

\bibitem{Komargodski:2017keh}
Z.~Komargodski and N.~Seiberg, {\it {A Symmetry Breaking Scenario for
  QCD$_3$}},  \href{http://xxx.lanl.gov/abs/1706.08755}{{\tt 1706.08755}}.

\bibitem{Son:2015xqa}
D.~T. Son, {\it {Is the Composite Fermion a Dirac Particle?}},  {\em Phys.
  Rev.} {\bf X5} (2015), no.~3 031027,
  \href{http://xxx.lanl.gov/abs/1502.03446}{{\tt 1502.03446}}.

\bibitem{ws}
C.~{Wang} and T.~{Senthil}, {\it {Dual Dirac Liquid on the Surface of the
  Electron Topological Insulator}},  {\em Physical Review X} {\bf 5} (Oct.,
  2015) 041031, \href{http://xxx.lanl.gov/abs/1505.05141}{{\tt 1505.05141}}.

\bibitem{mv}
M.~A. {Metlitski} and A.~{Vishwanath}, {\it {Particle-vortex duality of
  two-dimensional Dirac fermion from electric-magnetic duality of
  three-dimensional topological insulators}},  {\em Physical Review B} {\bf 93}
  (June, 2016) 245151, \href{http://xxx.lanl.gov/abs/1505.05142}{{\tt
  1505.05142}}.

\bibitem{mam}
D.~F. {Mross}, J.~{Alicea}, and O.~I. {Motrunich}, {\it {Explicit Derivation of
  Duality between a Free Dirac Cone and Quantum Electrodynamics in (2 +1 )
  Dimensions}},  {\em Physical Review Letters} {\bf 117} (July, 2016) 016802,
  \href{http://xxx.lanl.gov/abs/1510.08455}{{\tt 1510.08455}}.

\bibitem{Karch:2016sxi}
A.~Karch and D.~Tong, {\it {Particle-Vortex Duality from 3d Bosonization}},
  {\em Phys. Rev.} {\bf X6} (2016), no.~3 031043,
  \href{http://xxx.lanl.gov/abs/1606.01893}{{\tt 1606.01893}}.

\bibitem{Seiberg:2016gmd}
N.~Seiberg, T.~Senthil, C.~Wang, and E.~Witten, {\it {A Duality Web in 2+1
  Dimensions and Condensed Matter Physics}},  {\em Annals Phys.} {\bf 374}
  (2016) 395--433, \href{http://xxx.lanl.gov/abs/1606.01989}{{\tt 1606.01989}}.

\bibitem{Murugan:2016zal}
J.~Murugan and H.~Nastase, {\it {Particle-vortex duality in topological
  insulators and superconductors}},
  \href{http://xxx.lanl.gov/abs/1606.01912}{{\tt 1606.01912}}.

\bibitem{Karch:2016aux}
A.~Karch, B.~Robinson, and D.~Tong, {\it {More Abelian Dualities in 2+1
  Dimensions}},  \href{http://xxx.lanl.gov/abs/1609.04012}{{\tt 1609.04012}}.

\bibitem{Chen:2017lkr}
J.-Y. Chen, J.~H. Son, C.~Wang, and S.~Raghu, {\it {Exact Boson-Fermion Duality
  on a 3D Euclidean Lattice}},  \href{http://xxx.lanl.gov/abs/1705.05841}{{\tt
  1705.05841}}.

\bibitem{Aharony:2012ns}
O.~Aharony, S.~Giombi, G.~Gur-Ari, J.~Maldacena, and R.~Yacoby, {\it {The
  Thermal Free Energy in Large N Chern-Simons-Matter Theories}},  {\em JHEP}
  {\bf 03} (2013) 121, \href{http://xxx.lanl.gov/abs/1211.4843}{{\tt
  1211.4843}}.

\bibitem{Jain:2014nza}
S.~Jain, M.~Mandlik, S.~Minwalla, T.~Takimi, S.~R. Wadia, and S.~Yokoyama, {\it
  {Unitarity, Crossing Symmetry and Duality of the S-matrix in large N
  Chern-Simons theories with fundamental matter}},  {\em JHEP} {\bf 04} (2015)
  129, \href{http://xxx.lanl.gov/abs/1404.6373}{{\tt 1404.6373}}.

\bibitem{Inbasekar:2017ieo}
K.~Inbasekar, S.~Jain, P.~Nayak, and V.~Umesh, {\it {All tree level scattering
  amplitudes in Chern-Simons theories with fundamental matter}},
  \href{http://xxx.lanl.gov/abs/1710.04227}{{\tt 1710.04227}}.

\bibitem{Jensen:2017dso}
K.~Jensen and A.~Karch, {\it {Bosonizing three-dimensional quiver gauge
  theories}},  {\em JHEP} {\bf 11} (2017) 018,
  \href{http://xxx.lanl.gov/abs/1709.01083}{{\tt 1709.01083}}.

\bibitem{Jensen:2017xbs}
K.~Jensen and A.~Karch, {\it {Embedding three-dimensional bosonization
  dualities into string theory}},  {\em JHEP} {\bf 12} (2017) 031,
  \href{http://xxx.lanl.gov/abs/1709.07872}{{\tt 1709.07872}}.

\bibitem{Armoni:2017jkl}
A.~Armoni and V.~Niarchos, {\it {Phases of QCD$_3$ from Non-SUSY Seiberg
  Duality and Brane Dynamics}},  \href{http://xxx.lanl.gov/abs/1711.04832}{{\tt
  1711.04832}}.

\bibitem{Radicevic:2015yla}
D.~Radicevic, {\it {Disorder Operators in Chern-Simons-Fermion Theories}},
  {\em JHEP} {\bf 03} (2016) 131,
  \href{http://xxx.lanl.gov/abs/1511.01902}{{\tt 1511.01902}}.

\bibitem{Aharony:2016jvv}
O.~Aharony, F.~Benini, P.-S. Hsin, and N.~Seiberg, {\it {Chern-Simons-matter
  dualities with $SO$ and $USp$ gauge groups}},
  \href{http://xxx.lanl.gov/abs/1611.07874}{{\tt 1611.07874}}.

\bibitem{Metlitski:2016dht}
M.~A. Metlitski, A.~Vishwanath, and C.~Xu, {\it {Duality and bosonization of
  (2+1)d Majorana fermions}},  \href{http://xxx.lanl.gov/abs/1611.05049}{{\tt
  1611.05049}}.

\bibitem{Gaiotto:2017tne}
D.~Gaiotto, Z.~Komargodski, and N.~Seiberg, {\it {Time-Reversal Breaking in
  QCD$_4$, Walls, and Dualities in 2+1 Dimensions}},
  \href{http://xxx.lanl.gov/abs/1708.06806}{{\tt 1708.06806}}.

\bibitem{Gomis:2017ixy}
J.~Gomis, Z.~Komargodski, and N.~Seiberg, {\it {Phases Of Adjoint QCD$_3$ And
  Dualities}},  \href{http://xxx.lanl.gov/abs/1710.03258}{{\tt 1710.03258}}.

\bibitem{Cordova:2017vab}
C.~Cordova, P.-S. Hsin, and N.~Seiberg, {\it {Global Symmetries, Counterterms,
  and Duality in Chern-Simons Matter Theories with Orthogonal Gauge Groups}},
  \href{http://xxx.lanl.gov/abs/1711.10008}{{\tt 1711.10008}}.

\bibitem{Aitken:2017nfd}
K.~Aitken, A.~Baumgartner, A.~Karch, and B.~Robinson, {\it {3d Abelian
  Dualities with Boundaries}},  \href{http://xxx.lanl.gov/abs/1712.02801}{{\tt
  1712.02801}}.

\bibitem{Benini:2017dus}
F.~Benini, P.-S. Hsin, and N.~Seiberg, {\it {Comments on global symmetries,
  anomalies, and duality in (2 + 1)d}},  {\em JHEP} {\bf 04} (2017) 135,
  \href{http://xxx.lanl.gov/abs/1702.07035}{{\tt 1702.07035}}.

\bibitem{Jain:2013gza}
S.~Jain, S.~Minwalla, and S.~Yokoyama, {\it {Chern Simons duality with a
  fundamental boson and fermion}},  {\em JHEP} {\bf 11} (2013) 037,
  \href{http://xxx.lanl.gov/abs/1305.7235}{{\tt 1305.7235}}.

\bibitem{Gur-Ari:2015pca}
G.~Gur-Ari and R.~Yacoby, {\it {Three Dimensional Bosonization From
  Supersymmetry}},  {\em JHEP} {\bf 11} (2015) 013,
  \href{http://xxx.lanl.gov/abs/1507.04378}{{\tt 1507.04378}}.

\bibitem{Kachru:2016rui}
S.~Kachru, M.~Mulligan, G.~Torroba, and H.~Wang, {\it {Bosonization and Mirror
  Symmetry}},  {\em Phys. Rev.} {\bf D94} (2016), no.~8 085009,
  \href{http://xxx.lanl.gov/abs/1608.05077}{{\tt 1608.05077}}.

\bibitem{Kachru:2016aon}
S.~Kachru, M.~Mulligan, G.~Torroba, and H.~Wang, {\it {The many faces of mirror
  symmetry}},  \href{http://xxx.lanl.gov/abs/1609.02149}{{\tt 1609.02149}}.

\bibitem{Benini:2017aed}
F.~Benini, {\it {Three-dimensional dualities with bosons and fermions}},
  \href{http://xxx.lanl.gov/abs/1712.00020}{{\tt 1712.00020}}.

\bibitem{Shenker:2011zf}
S.~H. Shenker and X.~Yin, {\it {Vector Models in the Singlet Sector at Finite
  Temperature}},  \href{http://xxx.lanl.gov/abs/1109.3519}{{\tt 1109.3519}}.

\bibitem{Wu:1976ge}
T.~T. Wu and C.~N. Yang, {\it {Dirac Monopole Without Strings: Monopole
  Harmonics}},  {\em Nucl. Phys.} {\bf B107} (1976) 365.

\bibitem{Radicevic:2016wqn}
D.~Radicevic, D.~Tong, and C.~Turner, {\it {Non-Abelian 3d Bosonization and
  Quantum Hall States}},  {\em JHEP} {\bf 12} (2016) 067,
  \href{http://xxx.lanl.gov/abs/1608.04732}{{\tt 1608.04732}}.

\bibitem{Giveon:2008zn}
A.~Giveon and D.~Kutasov, {\it {Seiberg Duality in Chern-Simons Theory}},  {\em
  Nucl. Phys.} {\bf B812} (2009) 1--11,
  \href{http://xxx.lanl.gov/abs/0808.0360}{{\tt 0808.0360}}.

\bibitem{Intriligator:1996ex}
K.~A. Intriligator and N.~Seiberg, {\it {Mirror symmetry in three-dimensional
  gauge theories}},  {\em Phys. Lett.} {\bf B387} (1996) 513--519,
  \href{http://xxx.lanl.gov/abs/hep-th/9607207}{{\tt hep-th/9607207}}.

\bibitem{upcoming}
O.~Aharony, S.~Jain, and S.~Minwalla, {\it {Forthcoming}}, .

\bibitem{Benini:2011mf}
F.~Benini, C.~Closset, and S.~Cremonesi, {\it {Comments on 3d Seiberg-like
  dualities}},  {\em JHEP} {\bf 10} (2011) 075,
  \href{http://xxx.lanl.gov/abs/1108.5373}{{\tt 1108.5373}}.

\end{thebibliography}\endgroup

\end{document}